\pdfoutput=1
\documentclass[sigconf,authordraft=False,anonymous=False,review=False]{acmart}

\AtBeginDocument{%
  \providecommand\BibTeX{{%
    \normalfont B\kern-0.5em{\scshape i\kern-0.25em b}\kern-0.8em\TeX}}}

\copyrightyear{2023} 
\acmYear{2023} 
\setcopyright{acmlicensed}
\acmConference[WWW '23]{Proceedings of the ACM Web Conference 2023}{May 1--5, 2023}{Austin, TX, USA}
\acmBooktitle{Proceedings of the ACM Web Conference 2023 (WWW '23), May 1--5, 2023, Austin, TX, USA}
\acmPrice{15.00}
\acmDOI{10.1145/3543507.3583454}
\acmISBN{978-1-4503-9416-1/23/04}

\acmPrice{15.00}
\acmISBN{978-1-4503-XXXX-X/23/06}

\usepackage{array}
\usepackage{multirow}
\usepackage{multicol}
\usepackage{makecell}
\usepackage{bbm}
\usepackage{bm}
\usepackage{algorithm}
\usepackage{subfigure}

\usepackage{amsmath}

\begin{document}

\title{Homophily-oriented Heterogeneous Graph Rewiring}


\author{Jiayan Guo}
\authornote{Work performed during the internship at MSRA.}
\email{guojiayan@pku.edu.cn}
\affiliation{%
  \institution{School of Intelligence Science and Technology, Peking University}
  \state{Beijing}
  \country{China}
}

\author{Lun Du}
\authornote{Corresponding authors}
\email{lun.du@microsoft.com}
\affiliation{%
  \institution{Microsoft Research Asia}
  \city{Beijing}
  \country{China}
}

\author{Wendong Bi}
\authornotemark[1]
\email{biwendong20@mails.ucas.ac.cn}
\affiliation{%
  \institution{Institute of Computing Technology, Chinese Academy of Sciences}
  \city{Beijing}
  \country{China}
}

\author{Qiang Fu}
\email{qifu@microsoft.com}
\affiliation{
 \institution{Microsoft Research Asia}
 \state{Beijing}
 \country{China}
 }

\author{Xiaojun Ma}
\email{xiaojunma@microsoft.com}
\affiliation{%
  \institution{Microsoft Research Asia}
  \state{Beijing}
  \country{China}
  }

\author{Xu Chen}
\email{xu.chen@microsoft.com}
\affiliation{
  \institution{Microsoft Research Asia}
  \state{Beijing}
  \country{China}
}

\author{Shi Han}
\email{shihan@microsoft.com}
\affiliation{%
  \institution{Microsoft Research Asia}
  \city{Beijing}
  \country{China}
}

\author{Dongmei Zhang}
\email{dongmeiz@microsoft.com}
\affiliation{%
  \institution{Microsoft Research Asia}
  \city{Beijing}
  \country{China}
}

\author{Yan Zhang}
\email{zhyzhy001@pku.edu.cn}
\affiliation{
  \institution{School of Intelligence Science and Technology, Peking University}
  \city{Beijing}
  \country{China}
}

\renewcommand{\shortauthors}{Jiayan Guo and Lun Du, et al.}

\begin{abstract}

    With the rapid development of the World Wide Web~(WWW), heterogeneous graphs~(HG) have explosive growth. Recently, heterogeneous graph neural network~(HGNN) has shown great potential in learning on HG. Current studies of HGNN mainly focus on some HGs with strong homophily properties~(nodes connected by meta-path tend to have the same labels), while few discussions are made in those that are less homophilous. Recently, there have been many works on homogeneous graphs with heterophily. However, due to heterogeneity, it is non-trivial to extend their approach to deal with HGs with heterophily. In this work, based on empirical observations, we propose a meta-path-induced metric to measure the homophily degree of a HG. We also find that current HGNNs may have degenerated performance when handling HGs with less homophilous properties. Thus it is essential to increase the generalization ability of HGNNs on non-homophilous HGs. To this end, we propose HDHGR, a homophily-oriented deep heterogeneous graph rewiring approach that modifies the HG structure to increase the performance of HGNN. We theoretically verify HDHGR. In addition, experiments on real-world HGs demonstrate the effectiveness of HDHGR, which brings at most more than 10\% relative gain.


\end{abstract}




\maketitle

\section{Introduction}







Heterogeneous graph~(HG), also known as heterogeneous information network~\cite{sun2012mining}, has shown great potential in representing complex semantic relationships among real-world entities. Recently, heterogeneous graph neural network~(HGNN) arises and becomes a popular approach in learning on HG~\cite{wang2019heterogeneous,fu2020magnn,lv2021we,hu2020heterogeneous,yun2019graph,yu2022multiplex}. As a special kind of graph neural network~(GNN), it is able to capture attribute and semantic structural patterns of HG and has been widely used in many applications like recommender systems~\cite{10.1145/3488560.3498524,10.1145/3292500.3330673,hu-etal-2020-graph,chen2021graph,pang2022heterogeneous,10.1145/3477495.3532043,Guo2022EvolutionaryPL,Zhang2022EfficientlyLM}, drug discovery~\cite{yu2022hgdti,li2022predicting,zhao2019relation}, fraud detection~\cite{li2021live,wang2021modeling,wen2020asa,liu2021intention,liu2018heterogeneous} and natural language processing~\cite{linmei2019heterogeneous,jia2020neural,sheng2020summarize}. 


Homophily is a well-known property in real-world graphs~\cite{mcpherson2001birds,kossinets2009origins}. Recent stuties~\cite{zhu2020beyond,yan2021two,du2022gbk} designs ways to measure the homophily of a graph and find that the homophily property of a graph plays a vital role in influencing the performance of GNNs on homogeneous graphs. Current research on HGNN usually focus on some HGs with strong homophily property like academic collaboration network, i.e., authors of the same research direction usually publish papers at similar conferences. However, in many situations, HGs with less homophilous properties exist. For example, in a movie association network, movies performed by the same actor usually have different categories. However, due to the nature of heterogeneity, it is hard to directly extend the homophily measurement and the results derived from homogeneous graph to heterogeneous graph. In our research, we conduct empirical studies and find that the performance of HGNN is related to the homophily property of meta-path subgraphs. We find HGNNs tend to perform better on HGs with homophilous meta-path subgraphs while performing worse on HGs with only non-homophilous meta-path subgraphs. The observations show that HGNN tends to have degenerated performance on HG with less homophilous properties. Thus inspecting the homophily degree of a HG would be helpful when deploying HGNN models. However, 1) \textit{how to measure the degree of a HG} and 2) \textit{how to make current HGNNs generalize better on heterophily HGs} remains unresolved.

In this work, based on the observation that HGNN performance tends to be related to the most homophilous meta-path subgraph, we design a meta-path-induced metric to measure the homophily degree of a HG. Further, we design a homophily-oriented deep heterogeneous graph rewiring approach~(HDHGR) to modify the HG structure to increase the generalization ability of HGNN on non-homophilous HGs. Specifically, based on the intuition that nodes in the same class tend to have similar local distribution statistics, HDHGR uses local neighborhood feature distribution and label distribution to compute the similarities among node pairs. A meta-path similarity leaner~(MSL) is then used to learn pair-wise similarity under different meta-paths. It is trained via multi-objective optimization to balance the importance of different meta-paths. The learned MSL can be used to rewire meta-path subgraphs by removing edges with low similarities and adding edges with high similarities. The method is generic and can be applied to any kind of HGNN. For meta-path-based methods, e.g., HAN and MAGNN, the rewired meta-path instances can be used as their input.
Moreover, the rewired meta-path subgraphs can be seen as new auxiliary relations between target nodes for models that directly process the whole HG, e.g., GCN, GAT, and SHGN. In addition, from the complexity measurement perspective, we provide the theoretical analysis of the benefit of our approach for improving the generalization ability of HGNNs. The effectiveness of HDHGR is verified through empirical studies on non-homophilous HGs and homophilous HGs.

Our contributions are summarized as follows:

\begin{itemize}
    \item We design a meta-path-induced metric to measure the homophily degree of HGs. Such a metric enables us to do quantitative analysis and identify the insightful observations that homophily has a big impact on the performance of HGNN.
    \item We propose HDHGR, a homophily-oriented deep heterogeneous graph rewiring approach to boost the performance of HGNNs by reducing heterophily. The method is generic and can be applied to HGNNs with or without meta-path input. From the complexity measurement perspective, we provide the theoretical analysis of the benefit of our approach for improving the generalization ability of HGNNs.
    \item Extensive experiments on diverse HG datasets, including both non-homophilous HGs and homophilous HGs,  demonstrate the effectiveness of our method, which brings at most more than 10\% relative gain on non-homophilous HGs.
\end{itemize}

\begin{figure}
    \centering
    \includegraphics[width=.85\linewidth]{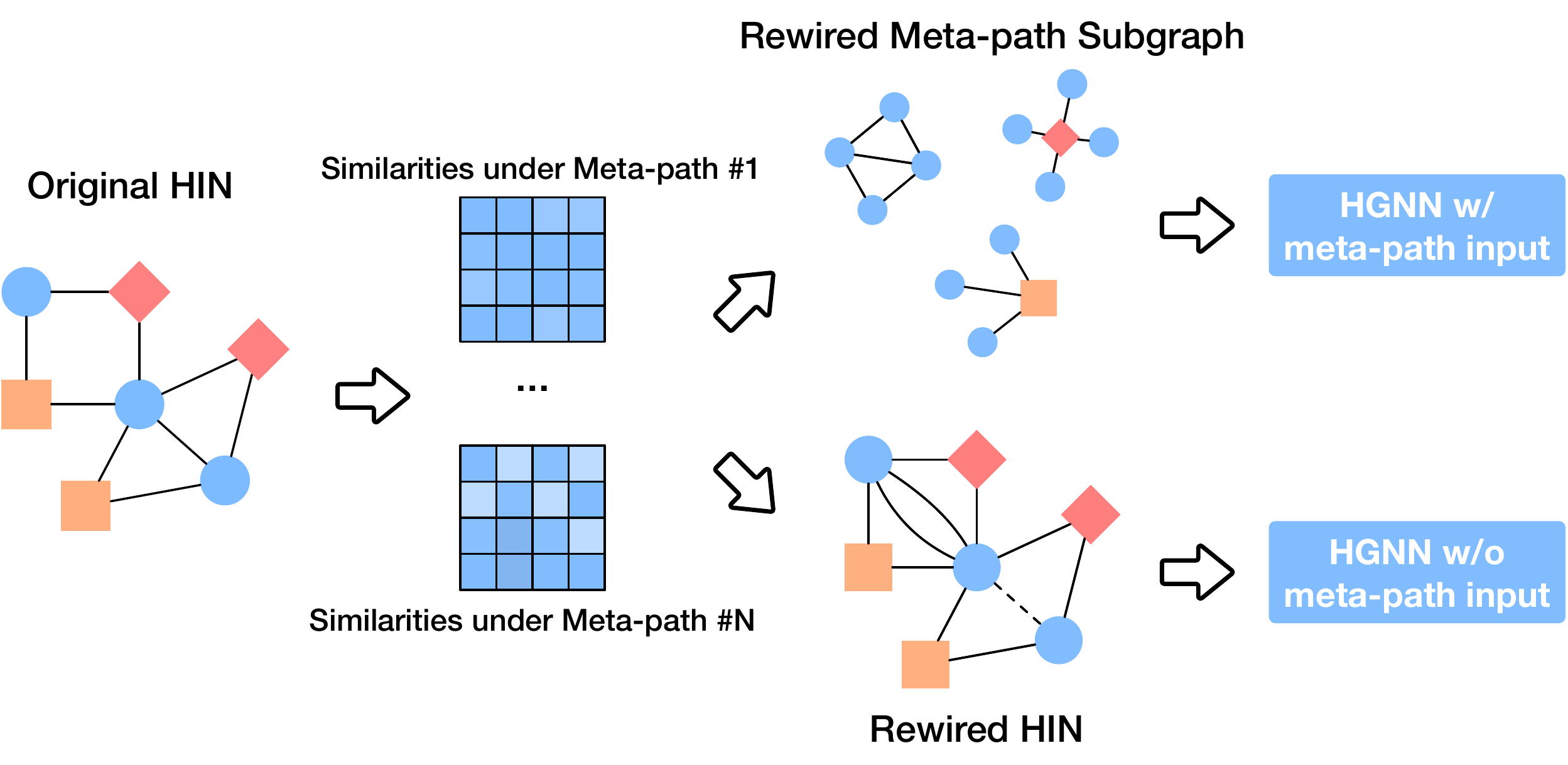}
    \caption{Pipeline of heterogeneous graph rewiring. It computes the similarity matrix under different meta-paths and uses it to rewire the HG for better performance on HGNN.}
    \label{fig:overview1}
\end{figure}

\section{Preliminary}

\noindent\textbf{Definition 2.1 Heterogeneous Graph.} Heterogeneous graph~(HG) can be represented as $\mathcal{G}=(\mathcal{V},\mathcal{E},\phi,\psi)$, where $\mathcal{V}={v_i}$ is the node set and $\mathcal{E}$ is the edge set. $\phi:\mathcal{V}\rightarrow\mathcal{T}$ maps nodes to their corresponding types, where $\mathcal{T}=\{\phi(v):v\in\mathcal{V}\}$. Similarly, $\psi:\mathcal{E}\rightarrow\mathcal{R}$ maps each edge to the type set, where $\mathcal{R}=\{\psi(e):e\in\mathcal{E}\}$.

\noindent\textbf{Definition 2.2 Meta-path.} Meta-path defines the semantic relationships between nodes in HG.  A meta-path P is defined as a path in the form of $\mathcal{T}_1\xrightarrow{R_1}\mathcal{T}_2\xrightarrow{R_2}\cdots\xrightarrow{R_L}\mathcal{T}_L$~(abbreviated as $\mathcal{T}_1\mathcal{T}_2\cdots\mathcal{T}_{L}$), which is the composition of the relation $R_1\circ R_2\circ\cdots\circ R_L$ between node type $\mathcal{T}_1$ and $\mathcal{T}_L$, where $\circ$ denotes the composition operator.

\noindent\textbf{Definition 2.3 Meta-path Subgraph.} A meta-path subgraph is a graph $G_\Phi=(\mathcal{V}_\Phi,\mathcal{E}_\Phi)$ induced by meta-path $\Phi=\mathcal{T}_1\cdots\mathcal{T}_L$. When $\mathcal{T}_1=\mathcal{T}_L$,  the meta-path subgraph is a homogeneous graph with edges in the relation defined by meta-path $\Phi$.


\noindent\textbf{Definition 2.4 Heterogeneous Graph Neural Network.} Heterogeneous graph neural networks typically map node features of different types into the same latent space via linear mapping. Then based on the mapped embeddings, one typical way is to leverage the message-passing mechanism to distinguish different relation types. Given self-loops a special type of edges, a classic HGNN like RGCN can be formulated as:
\begin{equation}
    \boldsymbol{h}^{(l+1)}_i=\sigma\left(\sum_{r\in\mathcal{R}}\sum_{j\in\mathcal{N}_r(i)}\frac{1}{\mathcal{N}_r(i)}\textbf{W}_r\boldsymbol{h}_{j}^{(l)}\right),
\end{equation}
\noindent where $\textbf{W}_r\in\mathbb{R}^{d\times d}$ is learnable weights. $\textbf{h}_j^{(l)}$ is the hidden state in layer-$l$. $\mathcal{N}_r(v_i)$ is the set of neighbors of node $v_i$ under relation $r$. HGCN captures information on various relations by assigning different weights to different edge types.

\noindent\textbf{Definition 2.5 Meta-path Subgraph Homophily Ratio.} Homophily ratio evaluates the overall homophily of a graph. It is defined as the ratio of edges connecting nodes with the same labels over the entire edge set:
\begin{equation}
    \label{eq:meta_path_homophiy_ratio}
    H(\mathcal{G}_\Phi)=\frac{\sum_{(v_i,v_j)\in\mathcal{E}_\Phi}\mathbbm{1}(y_i=y_j)}{|\mathcal{E}_\Phi|},
\end{equation}
\noindent where $\mathbbm{1}(\cdot)$ is the indicator function~(i.e., $\mathbbm{1}(\cdot)=1$ if the condition holds, otherwise $\mathbbm{1}(\cdot)=0$). $y_i$ is the label of node $v_i$ and $|\mathcal{E}|$ is the size of the edge set.

\noindent\textbf{Definition 2.6 Heterogeneous Graph Rewiring.} Given a heterogeneous graph $\mathcal{G} = (\mathcal{V}, \mathcal{E}, \phi, \psi)$ with node features $\textbf{X}\in\mathbb{R}^{N\times d}$ as the input, heterogeneous graph rewiring (MPSR) aims at learning an optimal graph $\mathcal{G}^* = (\mathcal{V}, \mathcal{E}^*,\phi,\psi)$ by updating the edge set with fixed node set under a specific criterion. Let $A,A^*\in\mathbb{R}^{N\times N}$  denotes the adjacent matrix of $\mathcal{G}$ and $\mathcal{G}^{*}$, respectively. The rewired heterogeneous graph $G^{*}$ is used as input of HGNNs, which is expected to be more effective than the original HG. As shown in Figure~\ref{fig:overview1}, the pipeline of heterogeneous graph rewiring models usually involves two stages: similarity learning and heterogeneous graph rewiring based on the learned similarity. 



\begin{figure*}
    \centering
    \includegraphics[width=.8\linewidth]{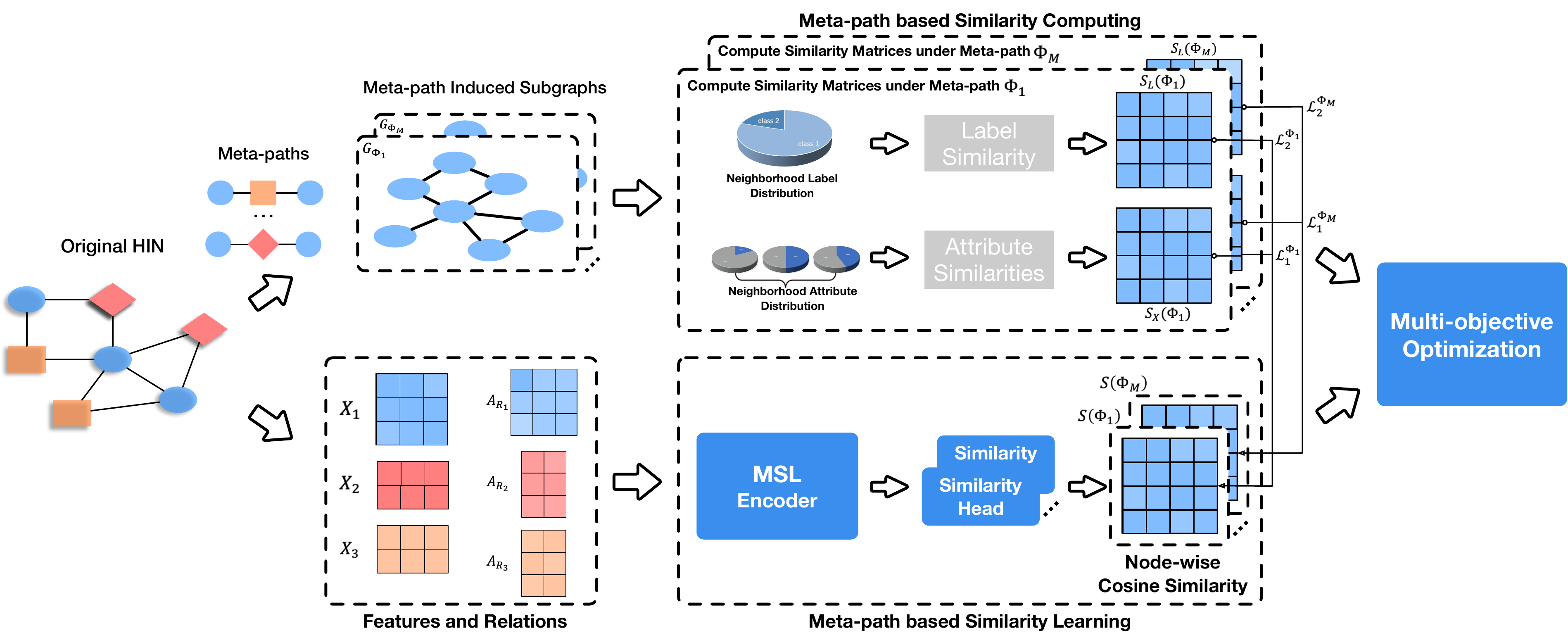}
    \caption{Overview of HDHGR. The meta-path subgraphs are first extracted and are used to compute the attribute similarity and label similarity matrices. Then the meta-path similarity learner~(MSL) takes the original HR as input and generates the similarity score under different semantic~(meta-path) space. Finally, the MSL is trained by multi-objective optimization.}
    \label{fig:overview}
\end{figure*}

\begin{table}[t]
    \centering
     \caption{Node classification result with GCN and HAN. HR denotes the meta-path subgraph homophily ratio in Sec 2.}
    \resizebox{.9\linewidth}{!}{
    \begin{tabular}{c|ccccc}
    \toprule
       Dataset & Meta-path & HR~(\%) & ACC~(\%) & \makecell[c]{Overall GCN\\ ACC~(\%)} & \makecell[c]{Overall HAN\\ ACC~(\%)} \\
    \midrule
       \multirow{5}{*}{ACM}  & PcP & 87.89 & $91.49\pm 0.27$ & \multirow{5}{*}{$92.12\pm 0.23$} & \multirow{5}{*}{$90.79\pm 0.43$}  \\
       & PAP & 79.36 & $89.76\pm 0.22$ & \\
       & PSP & 63.98 & $68.21\pm 1.13$ & \\
       & PcPAP & 75.84 & $87.54\pm 0.55$ & \\ 
       & PcPSP & 58.85 & $68.16\pm 1.22$ & \\
    \midrule
    \multirow{3}{*}{Liar} & NSpN & 21.25 & $24.75\pm 0.96$ & \multirow{3}{*}{$23.49\pm 0.57$} & \multirow{3}{*}{$26.75\pm 0.35$} \\
    &  NSuN   & 18.14 & $22.79\pm 0.41$ &  \\
    & NCN & 18.85 & $23.34\pm 0.67$ & \\
    \midrule
    \multirow{2}{*}{IMDB} & MDM & 61.41 & $52.78\pm 1.01$ & \multirow{2}{*}{$56.84\pm 2.66$} & \multirow{2}{*}{$55.81\pm 1.69$} \\
    & MAM & 44.43 & $51.31\pm 0.68$ & \\
    \bottomrule
    \end{tabular}}
    \label{tab:observation}
    \vspace{-0.2cm}
\end{table}

\section{Empirical Observations}
We conduct an empirical study to show the relation between meta-path subgraph homophily and HGNN performance. The results are reported in Table~\ref{tab:observation}. We use a 2-layer GCN with Relu activation and a hidden size of 64 as the base model for semi-supervised node classification task on each meta-path subgraph. We also train a GCN and a HAN on the entire HG following the settings of ~\cite{lv2021we}. From Table~\ref{tab:observation}, we have the following discoveries: 

\begin{enumerate}
    \item GCN consistently performs better on meta-path subgraph with a higher homophily ratio. In ACM, the highest homophily meta-path subgraph is induced by PcP, and the model achieves the best performance under this scene. In Liar and IMDB, the model also obtains the best classification results on meta-path subgraphs with relatively high homophily ratios.
    \item  Performance on the entire HG is related to the meta-path subgraph homophily ratio. The accuracy of GCN on the entire ACM graph is 92.12\%, which is almost the same as the result on PcP-induced meta-path subgraph. The evaluation metric in Liar and IMDB are 23.49\% and 56.84\%, which are also similar to the result of NSpN and MDM respectively.
\end{enumerate}

\noindent The two discoveries show that the homophily ratio of the meta-path-induced subgraph influences the performance of GCN. In Section~\ref{sec:exp} we further show that this conclusion holds for most HGNNs. Such results inspire us that we can improve the meta-path subgraph homophily ratio of HG to enable a better matching with HGNNs, and it leads to our key idea, a homophily-oriented heterogeneous graph rewiring method.

\noindent\textbf{Definition 3.1 Heterogeneous Graph Homophily Ratio.} Given a HG $\mathcal{G}=(\mathcal{V},\mathcal{E},\phi,\psi)$, and $k$-hop meta-path set $\mathcal{M}_k$, the heterogeneous graph homophily ratio of HG is defined by:
\begin{equation}
   MH(\mathcal{G})=\text{max}\left(\left\{H(\mathcal{G}_\Phi)|\Phi\in\mathcal{M}_k,k=1,...\right\}\right).
\end{equation}
It calculates the maximum 
 homophily ratio over subgraphs with different hops of meta-path, which is the maximum potential homophily ratio of a given graph.



\section{Methodology}

In this section, based on the observation and discussion above, we introduce the \textit{\underline{H}omophily-oriented \underline{D}eep \underline{H}eterogeneous \underline{G}raph \underline{R}ewiring} method~(HDHGR). We first show how to compute the meta-path-based similarity from both attribute and label perspectives. Then we give the description of the heterogeneous graph similarity learner and how to rewire the HG via the learner.   

\subsection{Meta-path based Similarity Computing}

Before training the similarity learner, we first construct the objective of the learner. We utilize two important factors that are helpful in preserving node similarities under the corresponding semantic space, 1) \textbf{neighborhood attribute distribution}~(NAD) and 2) \textbf{neighborhood label distribution}~(NLD). For a meta-path subgraph $\mathcal{G}_\Phi=(\mathcal{V}_\Phi,\mathcal{E}_\Phi)$, the $k$-hop NAD and $k$-hop NLD are computed as:
\begin{equation}
    D_\textbf{X}^{(k)}(\Phi)=(D_\Phi^{-1}A_\Phi)^k\textbf{X}_\tau \ , \ \ D_\textbf{Y}^{(k)}(\Phi)=(D_\Phi^{-1}A_\Phi)^k\textbf{Y}^{\text{train}}_\tau
\end{equation}
\noindent where $D_\textbf{X}^{(k)}\in\mathbb{R}^{N_\tau\times d_{\text{in}}}$ and $D_\textbf{Y}^{(k)}\in\mathbb{R}^{N_\tau\times c}$ are the $k$-hop attribute distribution and label distribution matrix of meta-path induced subgraph $G_\Phi$ respectively. $A_\Phi\in\mathbb{R}^{N_\tau\times N_\tau}$ is the adjacency matrix of $G_\Phi$ and $D_\Phi\in\mathbb{R}^{N_\tau\times N_\tau}$ is the corresponding degree diagonal matrix. $\textbf{X}_\tau\in\mathbb{R}^{N_\tau\times d_{\text{in}}}$ is the attribute matrix and $\textbf{Y}_\tau^{\text{train}}\in\mathbb{R}^{N_\tau\times c}$ is the one-hot label matrix of node type $\tau$ in the training set. The $i$-th row of $\textbf{Y}_\tau^{\text{train}}$ is a one-hot label vector if $v_i$ belongs to the training set else an all 0 vector. Then for each node, we can get the observed label-distribution vector and feature-distribution vector of its neighbors. Next we calculate the cosine distance between each node-pair in $G_\Phi$ with respect to $D_X^{(k)}$ and $D_Y^{(k)}$, which forms the attribute similarity matrix $S_X(\Phi)$ and the label similarity matrix $S_Y^{\text{train}}(\Phi)$:
\begin{equation}
\small
    \begin{cases}
    S_\textbf{X}(\Phi)[i,j]&=\prod_{k=1}^{K}\text{Cos}\left( D_\textbf{X}^{(k)}(\Phi)[i, :], D_\textbf{X}^{(k)}(\Phi)[j, :] \right) \\
    
    S_\textbf{Y}^{\text{train}}(\Phi)[i,j]&=\prod_{k=1}^{K}\text{Cos}\left( D_\textbf{Y}^{(k)}(\Phi)[i, :], D_\textbf{Y}^{(k)}(\Phi)[j, :] \right) \\
    \end{cases}
\end{equation}
\noindent where
\begin{equation}
    \text{Cos}(\boldsymbol{x}_i,\boldsymbol{x}_j)=\frac{\overline{\boldsymbol{x}}_i\cdot\overline{\boldsymbol{x}}_j}{\Vert\overline{\boldsymbol{x}}_i\Vert\Vert\overline{\boldsymbol{x}}_j\Vert} \ , \ \ \overline{\boldsymbol{x}}_i=\boldsymbol{x}_i-\frac{1}{|V_\tau|}\sum_{v_j\in V_\tau} x_j 
\end{equation}

The cosine similarity is calculated after subtracting the mean of the variable of all the nodes. It should be noted that we only use the labels in the training set, thus not all labels have observable values. To make the label similarity estimation more accurate, we only use the label distribution of a node when the portion of its neighbors that have observable labels exceeds a threshold. In specifically, we restrict the utilization condition of neighbor label-distribution by using a mask, by:
\begin{equation}
    \textbf{M}_\textbf{Y}^{\Phi}(v_i)=\begin{cases}
    1, &  \ r_i>\alpha \\
    0, &  \ r_i \le\alpha
    \end{cases}, \text{where $r_i=\frac{|\mathcal{N}_\Phi(v_i)\cap\mathcal{V}_\tau^{\text{train}}|}{\mathcal{N}_\Phi(v_i)}$} 
\end{equation}
\noindent where $\textbf{M}_\textbf{Y}^{\Phi}(v_i)$ is the mask vector, $\mathcal{N}_\Phi(v_i)$ is the neighborhood set of $v_i$ under meta-path $\Phi$, $\mathcal{V}_\tau^{\text{train}}$ is the set of nodes in the training set. We will show in detail how this mask can be used to filter out nodes with few label observable neighbors.

\subsection{Meta-path based Similarity Learning}

After constructing the training objective, we use a Meta-path Similarity Learner~(MSL) to fit the object. The MSL should consider the whole information of the original HG, thus we use a heterogeneous graph encoder to encode the whole graph. We first map each node type into the same hidden space by a linear mapping. To stay efficient, we take the HG as its homogeneous counterpart and use it to encode the graph, by:
\begin{equation}
    \textbf{H}^{(k)}=(D^{-1}A\textbf{H}^{(0)})^{k}\textbf{W}^{(l)}
\end{equation}
\noindent where $\textbf{H}_0$ is the mapped hidden states of nodes. As each meta-path contains different semantics, we use a different linear mapping encoder to map the node hidden states $\textbf{H}\in\mathbb{R}^{N_\tau\times d}$ into the corresponding space $\textbf{H}_\Phi\in\mathbb{R}^{N_\tau\times d}$:
\begin{equation}
    \textbf{H}_\Phi=\textbf{H}^{(k)}\textbf{W}_\Phi
\end{equation}

\noindent where $\textbf{W}_\Phi$ is the parameter under meta-path $\Phi$. Then we can use the representations to compute the semantic level similarities of each node pairs in meta-path induced subgraph $G_\Phi$:
\begin{equation}
    S(\Phi)[i,j]=\prod_{k=1}^{K}\text{Cos}(\textbf{h}_i^{(k)},\textbf{h}_j^{(k)})
\end{equation}
\noindent where $S(\Phi)[i,j]$ is the similarity score of node $v_i$ and $v_j$ under meta-path $\Phi$. In practice, we also optionally use the concatenation of distribution feature $\boldsymbol{h}_i^k$  and $\boldsymbol{h}_i\cdot \textbf{W}_\Phi$ for similarity calculation. We then use $S_\textbf{X}(\Phi)$ and $S_\textbf{Y}^{\text{train}}(\Phi)$ to guide the training of the MSL, by
\begin{equation}
    \begin{split}
        \mathcal{L}_1^{\Phi}&=\Vert S(\Phi)-S_\textbf{X}(\Phi)\Vert_F^2 \\
        \mathcal{L}_2^{\Phi}=\Vert (S(\Phi)&-S_\textbf{Y}(\Phi)^{\text{train}})\odot (\textbf{M}^{\Phi}_{\textbf{Y}} \textbf{M}^{{\Phi}^T}_{\textbf{Y}}) \Vert_F^2
    \end{split}
\end{equation}

The final loss function for meta-path $\Phi$ is the summation of the two loss functions:
\begin{equation}
    \mathcal{L}^{\Phi}=\mathcal{L}_1^{\Phi}+\mathcal{L}_2^{\Phi}
\end{equation}

\subsection{Multi-objective Optimization}

As each meta-path contains different semantic meanings and different similarities and may contain useful information for node relations, we need to optimize for all of them. However, directly optimizing each meta-path equally may make MSL unable to fully learn the knowledge of some important meta-paths. Besides, some meta-path may have a negative influence on the training of MSL. Thus we use multi-task learning and assign a unique weight $\lambda_\Phi$ for each meta-path to balance different objectives:
\begin{equation}
\small
    \mathcal{L}=\frac{1}{M}\sum_{i=1}^M\lambda_{\Phi_i}\mathcal{L}^{\Phi_k}
\end{equation}
\noindent where $\sum_{i=1}^M\lambda_{\Phi_i}=1$. We then use multi-objective optimization~\cite{sener2018multi} to determine the weight for each objective. In specific, it uses an efficient Frank-Wolfe-bsased optimizer to find the Pareto optimal solution of $\lambda_{\Phi_i},\ 
i=1\ldots,M$ under realistic assumptions. 

\subsection{Efficient Implementation}

Directly optimizing the objective function mentioned above has quadratic computational complexity. For node attributes $\textbf{X}_\tau\in\mathbb{R}^{N_\tau\times d}$ , the $O(N_\tau^2)$ complexity is unacceptable for large graphs when $N_\tau >> d$. So we design a scalable training strategy with stochastic mini-batch. Specifically, we randomly select $k_1\times k_2 $ target node-pairs as a batch and optimize the similarity matrix $S(\Phi)$ by a ($k_1\times k_2$ )-sized sliding window in each iteration. We can assign small numbers to $k_1, k_2\in[1, N_\tau]$. Besides, we can use the same batch to optimize different meta-paths at the same time.

\subsection{Meta-Path Subgraph Rewiring via \\ Similarity Learner}

After obtaining the similarity of each target node pair, we can use the learned similarity $S(\Phi)$ to rewire the meta-path $\Phi$. Specifically, we add edges between node pairs with high similarity and remove edges with low similarity on the original meta-path-induced subgraph. There are three parameters are set to control this process: $K$ determines the maximum number of edges that can be added for each node; 
$\epsilon$ 
constrains the threshold that the similarity of node-pairs to add edges must be larger. Another parameter $\gamma$ is set for pruning edges with a similarity smaller than $\gamma$.  Finally, for HGNN directly use meta-paths, we use the rewired meta-path instances as input. While for HGNN that does not need meta-paths, we add the rewired instances to the original HG $\mathcal{G}$. Then we can use these models to do node classification tasks.

\subsection{Complexity Analysis}

In this section, we analyze the computational complexity of MSC and MSL. For MSL, Given the hidden dimension for computing similarity matrix $D$, the number of hops is $K$. By using sparse matrix multiplication when computing multi-hop representations, constructing the feature and label similarity matrix takes $\mathcal{O}(K|\mathcal{V}||\mathcal{E}|+Dk_1k_2)$, where $k_1$ and $k_2$ are constants representing the batch size of target nodes. For MSL, the complexity is the same as MSC. Then the final time complexity is $\mathcal{O}(K|\mathcal{V}||\mathcal{E}|+Dk_1k_2)$. For space complexity, the HDHGR needs to store the computed attribute level and label level similarity matrices under each meta-path of target nodes thus are $\mathcal{O}(\mathcal{V}_\tau^2)$. Besides, it needs to store the parameters of MSL, which is $\mathcal{O}(K\mathcal{D}^2)$. The final space complexity is $\mathcal{O}(\mathcal{V}_\tau^2+K\mathcal{D}^2)$.

\begin{table}[t]
\Huge
    \centering
    \caption{Dataset Statistics.}
    \resizebox{.9\linewidth}{!}{
    \begin{tabular}{cccccccc}
    \toprule
      Dataset & \#Nodes & \makecell[c]{\#Nodes \\ Types} & \#Edges & \makecell[c]{\#Edge \\ Types} & Target & \#Classes & $MH$  \\
    \midrule
       FB-American  &  6,386 & 7 & 519,631 & 12 & person & 2 & 0.52 \\
       Actor & 16,255 & 3 & 99,265 & 18 & actor & 3 & 0.34 \\
       Liar & 14,395 & 4 & 146,935 & 6 & news & 7 & 0.21 \\
       IMDB & 11,616 & 3 & 17106 & 2 & movie & 3 & 0.61 \\
       ACM & 10,942 & 4 & 547,872 & 8 & paper & 3 & 0.88 \\
       DBLP & 26,128 & 4 & 239,566 & 6 & author & 4 & 0.80 \\
    \bottomrule
    \end{tabular}
    }
    \label{tab:statistics}
    \vspace{-0.3cm}
\end{table}

\section{Theoretical Analysis}

We theoretically analyze the role of homophily for HGNNs with complexity measure~(CM). The complexity measure is given by:

\noindent\textbf{Definition 1}
\textit{Complexity measure (CM) is a measure function $\mathcal{M}:{\mathcal{H},\mathcal{S}}\rightarrow\mathbb{R}^+$ to measure the complexity of models, where $\mathcal{H}$ is a class of models and $\mathcal{S}$ is a training set. \textbf{Note that a lower complexity measure means  better generalization ability.}}

Under the setting of the heterogeneous graph, we can set $\mathcal{H}$ as RGCN with different parameters and $\mathcal{S}$ as HG data. In this paper, we use Consistency of Representations~\cite{natekar2020representation} as the CM. This measure is designed based on Davies-Bouldin Index~\cite{Davies1979ACS}. Mathematically, for a given dataset and a given layer of a model,
\begin{equation}
\small
    S_i=\left(\frac{1}{n_k}\sum_{t=1}^{n_i}\vert\mathcal{O}_i^{(t)}-\mu_{\mathcal{O}_i}\vert^p\right)^{\frac{1}{p}}, \ \ i=1,...,k
\end{equation}
\begin{equation}
\small
    M_{i,j}=\vert\vert\mu_{\mathcal{O}_i}-\mu_{\mathcal{O}_j}\vert\vert_p, \ \ i,j=1,...,k
\end{equation}
\noindent where $i,j$ are the class index, $\mathcal{O}_{i}^{(t)}$ is the representation of $t$-th sample in class $i$, $\mu_{\mathcal{O}_{i}}$ is the cluster centroid of class $i$. $S_i$ can be seen as a scalar measure of the intra-class distance and $M_{i,j}$ can be seen as a scalar that measures the separation of representations of class $i$ and $j$. The Davies Bouldin Index CM can be written as:
\begin{equation}
\small
    \mathcal{C}=\frac{1}{k}\sum_{i=0}^{k-1}\mathop{\text{max}}_{i\neq j}\frac{S_i+S_j}{M_{i,j}}
\end{equation}
\noindent When $p=2$, the measure can be seen as the ratio of intra-class variance and inter-class variance.
Under these settings, we have the following theorem.

\noindent\textbf{Theorem 1.} 
\textit{
On the heterogeneous graph $\mathcal{G}=(\mathcal{V},\mathcal{E},\phi,\psi)$ with meta-path set $\mathcal{R}$, we consider a binary classification task under the condition that the neighbors of each node $v_i\in \mathcal{V}$ have the probability of $P^r$ to belong to the same class with $v_i$ under meta-path $r$. 
\textbf{For an arbitrary meta-path type $r\in\mathcal{R}$, when $P^r\rightarrow 1$, the  upper bound of the HGNN's generalization ability reaches optimal.}
}

And this theorem motivates us to improve the generalization ability of HGNN by reducing the heterophily degree of heterogeneous graphs. And in this paper, we propose to reduce the heterophily degree of the heterogeneous graph with graph rewiring. The whole proof of the theorem can be found in Appendix~\ref{app:theory}.

\section{Experiment}
\label{sec:exp}

\subsection{Experimental Setup}

\subsubsection{\textbf{Datasets}} We consider different kinds of HGs to demonstrate the effectiveness of our rewiring method. They are 4 heterophily HGs and 2 Homophily HGs:

Non-homophilous heterogeneous graphs:

\begin{itemize}
    \item \textbf{FB-American}\footnote{https://archive.org/details/oxford-2005-facebook-matrix}~\cite{traud2012social} One of the FB100 dataset. The dataset contains Facebook users in American University. It has 7 node types and 12 edge types. We use the dataset to predict the gender of the user. 
    \item \textbf{Actor}\footnote{https://lfs.aminer.cn/lab-datasets/soinf/}~\cite{tang2009social} A heterogeneous network data that crawled on Wiki. There are three kinds of nodes, actor, director and writer and 18 edge types. The task is to predict the actor's property based on the clusters of the actor description.
    \item \textbf{Liar}\footnote{https://huggingface.co/datasets/liar}~\cite{wang2017liar} A heterogeneous network data that is mainly used for fraud detection. The nodes contain news, speaker, Twitter, and context.  
    \item \textbf{IMDB}\footnote{https://www.imdb.com} is an online database about movies and television programs, including information such as cast, production crew, and plot summaries. We choose the version in~\cite{fu2020magnn} for fair comparisons.
\end{itemize}

homophilous heterogeneous graphs\footnote{https://github.com/THUDM/HGB}~\cite{lv2021we}:

\begin{itemize}
    \item \textbf{ACM} a citation network in heterogeneous graph benchmark. HGB uses the subset hosted in HAN~\cite{wang2019heterogeneous}, but preserves all edges including paper citations and references.
    \item \textbf{DBLP} is a bibliography website of computer science that is also in heterogeneous graph benchmark. HGB uses a commonly used subset in 4 areas with nodes representing authors, papers, terms, and venues.
\end{itemize}

\begin{table*}[t]
\Large
    \centering 
    \caption{Experimental Results on Non-homophilous HGs.}
    \resizebox{0.8\linewidth}{!}{
    \begin{tabular}{cccccccccc}
    \toprule
      \multirow{2}{*}{Model}  &  \multirow{2}{*}{/}  & \multicolumn{2}{c}{FB-American} & \multicolumn{2}{c}{Actor} & \multicolumn{2}{c}{Liar} & \multicolumn{2}{c}{IMDB} \\
      
          \cline{3-10}
       \multicolumn{2}{c}{} & Macro-F1 & Micro-F1 & Macro-F1 & Micro-F1 & Macro-F1 & Micro-F1 & Macro-F1 & Micro-F1 \\
         \midrule
        \multirow{2}{*}{GCN} & origin & $69.60\pm 0.88$ & $72.73\pm 0.76$ & $55.38\pm 0.32$ & $65.79\pm 0.21$ & $19.94\pm 1.83$ & $23.49\pm 0.57$ & $55.49\pm 4.28$ & $56.84\pm 2.66$\\
         & HDHGR & $\textbf{72.04}\pm \textbf{0.91}$ & $\textbf{74.70}\pm \textbf{0.89}$ & $\textbf{67.21}\pm \textbf{0.94}$ & $\textbf{75.07}\pm \textbf{0.38}$ & $\textbf{21.02}\pm \textbf{1.24}$ & $\textbf{24.08}\pm \textbf{0.37}$ & $\textbf{58.22}\pm \textbf{0.46}$ & $\textbf{58.97}\pm \textbf{0.41}$ \\
         \midrule
         \multirow{2}{*}{GAT} & origin & $70.91\pm 1.63$ & $73.83\pm 1.05$ & $45.78\pm 1.00$ & $60.91\pm 0.58$ & $20.35\pm 0.42$ & $24.12\pm 1.30 $ & $54.99\pm 3.16 $ & $55.82\pm 2.11$\\
         & HDHGR & $\textbf{73.23}\pm \textbf{2.25}$ & $\textbf{75.53}\pm \textbf{1.82}$ & $\textbf{63.83}\pm \textbf{0.56}$ & $\textbf{74.23}\pm \textbf{0.28}$ & $\textbf{21.50}\pm \textbf{0.64}$ & $\textbf{24.67}\pm \textbf{0.70}$ & $\textbf{57.26}\pm \textbf{3.10}$ & $\textbf{57.99}\pm \textbf{2.16}$ \\
         \midrule
         \multirow{2}{*}{$\text{H}_2\text{GCN}$} & origin & $75.97\pm 0.47$ & $77.58\pm 0.36$ & $49.70\pm 0.67$ & $63.65\pm 0.76$ & $19.80\pm 0.57$ & $23.23\pm 0.48$ & $53.44\pm 0.68$ & $54.29\pm 0.62$\\
         & HDHGR & $\textbf{76.44}\pm \textbf{0.68}$ & $\textbf{78.30}\pm \textbf{0.64}$ & $\textbf{62.67}\pm \textbf{1.13}$ & $\textbf{72.66}\pm \textbf{0.51}$ & $\textbf{22.71}\pm \textbf{0.80}$ & $\textbf{23.27}\pm \textbf{0.13}$ & $\textbf{55.47}\pm \textbf{0.24}$ & $\textbf{56.03}\pm \textbf{0.24}$ \\
         \midrule
         \multirow{2}{*}{LINKX} & origin & $72.85\pm 0.23$ & $73.05\pm 0.24$ & $57.98\pm 2.01$ & $63.97\pm 1.77$ & $15.02\pm 0.25$ & $20.78\pm 0.42$ & $43.88\pm 1.22$ & $44.11\pm 1.12$  \\
         & HDHGR & $\textbf{73.73}\pm \textbf{0.52}$ & $\textbf{74.35}\pm \textbf{0.49}$ & $\textbf{63.79}\pm \textbf{2.05}$ & $\textbf{70.23}\pm \textbf{1.83}$ & $\textbf{23.09}\pm \textbf{0.23}$ & $\textbf{24.20}\pm \textbf{0.33}$ & $\textbf{47.99}\pm \textbf{0.78}$ & $\textbf{48.30}\pm \textbf{0.85}$\\
         \midrule
         \multirow{2}{*}{RGCN} & origin & $51.50\pm 1.14 $ & $63.88\pm 0.17$ & $49.21\pm 2.48$ & $71.61\pm 0.82$ & $17.87\pm 0.51$ & $23.54\pm 0.10$ & $50.33\pm 4.40$ & $52.51\pm 2.67$  \\
         & HDHGR & $\textbf{55.50}\pm \textbf{1.20}$ & $\textbf{65.69}\pm \textbf{0.17}$ & $\textbf{53.33}\pm \textbf{1.58}$ & $\textbf{74.83}\pm \textbf{0.39}$ & $\textbf{19.55}\pm \textbf{0.10}$ & $\textbf{24.69}\pm \textbf{0.14}$ & $\textbf{55.07}\pm \textbf{1.08}$ & $\textbf{55.45}\pm \textbf{0.71}$  \\
         \midrule
         \multirow{2}{*}{HAN} & origin & $58.86\pm 0.83$ & $66.28\pm 0.60$ & $54.78\pm 1.54$ & $68.95\pm 0.26$ & $22.51\pm 0.90$ & $26.75\pm 0.35$& $53.85\pm 3.31$ & $55.81\pm 1.69$ \\
         & HDHGR & $\textbf{64.93}\pm \textbf{1.87}$ & $\textbf{69.99}\pm \textbf{0.96}$ & $\textbf{71.13}\pm \textbf{1.38}$ & $\textbf{80.89}\pm \textbf{0.39}$ & $\textbf{22.87}\pm \textbf{1.08}$ & $\textbf{27.33}\pm \textbf{0.37}$ & $\textbf{58.97}\pm \textbf{0.58}$ & $\textbf{59.32}\pm \textbf{0.53}$\\
          \midrule
          \multirow{2}{*}{HGT} & origin & $55.65\pm 0.50$ & $64.60\pm 0.41$ & $75.58\pm 0.48$ & $82.97\pm 0.29$ & $21.38\pm 0.61$ & $26.27\pm 0.55$ & $39.34\pm 0.54$ & $50.88\pm 0.41$ \\
          &  HDHGR  & $\textbf{62.49}\pm \textbf{1.37}$ & $\textbf{67.43}\pm \textbf{1.41}$ & $\textbf{77.25}\pm \textbf{0.53}$ & $\textbf{84.06}\pm \textbf{0.29}$ & $\textbf{21.87}\pm \textbf{0.54}$ & $\textbf{26.66}\pm \textbf{0.37}$ & $\textbf{49.46}\pm \textbf{2.67}$ & $\textbf{52.93}\pm \textbf{1.62}$ \\
          \midrule
         \multirow{2}{*}{SHGN} & origin & $74.62\pm 4.30$ & $76.72\pm 3.76$ & $71.66\pm 1.62$ & $80.85\pm 0.29$ & $18.45\pm 1.60$ & $23.45\pm 0.76$ & $51.60\pm 1.07$ & $52.50\pm 0.91$  \\
         & HDHGR & $\textbf{76.03}\pm \textbf{1.65}$ & $\textbf{77.84}\pm \textbf{1.50}$ & $\textbf{72.73}\pm \textbf{0.35}$ & $\textbf{81.36}\pm \textbf{0.37}$ & $\textbf{22.29}\pm \textbf{0.72}$ & $\textbf{23.82}\pm \textbf{1.08}$  & $\textbf{55.69}\pm \textbf{0.61} $ & $\textbf{56.56}\pm \textbf{0.48}$ \\
         \midrule
         ARI & / & $5.11\%$ & $2.75\%$ & $17.37\%$ & $10.46\%$ & $12.93\%$ & $3.99\%$ & $9.35\%$ & $5.50\%$ \\
         \bottomrule
    \end{tabular}
    }
    \label{tab:main_exp}
\end{table*}

\subsubsection{\textbf{Baselines}} HDHGR can be decoupled with HGNN training. Then we select 4 homogeneous GNNs and 5 heterogeneous GNNs. Besides, to validate the effectiveness of  HDHGR as a heterogeneous graph rewiring method, we also compare DHGR with three heterogeneous graph structure learning methods.

Homogeneous graph neural networks:

\begin{itemize}
    \item\textbf{GCN}~\cite{kipf2016semi}  is a semi-supervised graph convolutional network model which learns node representations by aggregating information from neighbors.
    \item\textbf{GAT}~\cite{velivckovic2017graph}  is a graph neural network model using the attention mechanism to aggregate node features.
    \item$\textbf{H}_{2}\textbf{GCN}$~\cite{zhu2020beyond} is a graph neural network that combines ego and neighbor-embedding separation, higher-order neighborhoods, and a combination of intermediate representations, which performs better on heterophily graph data.
    \item\textbf{LINKX}~\cite{lim2021large} is a graph neural network that decouples structure and feature transformation and combines both information. It performs better on heterophily graph data.
\end{itemize}

Heterogeneous graph neural networks:

\begin{itemize}
    \item\textbf{RGCN}~\cite{schlichtkrull2018modeling} is a HGNN that assigns different parameters on different relations, and representations of neighbors are aggregated under different relations.
    \item\textbf{HAN}~\cite{wang2019heterogeneous} is an attention-based HGNN that designs instance-level and semantic-level attention to aggregate neighborhood information along different meta-paths.
    \item\textbf{HGT}~\cite{hu2020heterogeneous} a self-attention-based HGNN that assigns learnable weight for each node type and relations. 
    \item\textbf{SHGN}~\cite{lv2021we} a HGNN that distinguishes different relations when computing attention score and uses $l_2$-norm to normalize the model output.
\end{itemize}

Graph structure learning and rewiring methods:

\begin{itemize}
    \item\textbf{LDS}~\cite{franceschi2019learning} is a graph structure learning method that uses bi-level optimization to optimize the graph structure learning. 
    \item\textbf{IDGL}~\cite{chen2020iterative} is a graph structure learning method that iteratively updates the learned graph structure by the guidance of the task performance.
    
    \item\textbf{HGSL}~\cite{zhao2021heterogeneous} is a heterogeneous graph structure learning method that designs attribute, meta-path, and structural representations to guide relation learning.
    
\end{itemize}

\subsubsection{\textbf{Implementation}} For baseline GNN methods, we use the best hyper-parameter reported in their original paper. For HGNN baselines, we use the implementation released by HGB~\cite{lv2021we} and tune through the best effort. For graph structure learning methods that are originally on homogeneous graphs, we ignore the node and edge type of the HG and use it as the input. For each HG dataset, following the typical processing of previous works~\cite{wang2019heterogeneous,fu2020magnn}, we choose one node type as a target with training/validation/test splitting ratio 0.6/0.2/0.2. For HDHGR, we perform grid search to choose the best hyper-parameters. The growing threshold $K$ is searched in $\{3,6,8,16\}$. The minimum similarity threshold $\epsilon$ is searched in \{0.0, 0.3,0.6,0.9\}. The degree threshold $\alpha$ is searched in $\{0.1,0.3,0.6,0.9,0.95\}$. We train HDHGR 200 epochs with $S_\textbf{X}(\Phi)$ and finetune 30 epochs with $S^{\text{train}}_\textbf{Y}(\Phi)$. The result is averaged under 5 random seeds. We use Adam optimizer to optimize all methods. The learning rate is fixed to $5e{-4}$ and the weight decay is $1e{-4}$. We use 4 Tesla V100 GPUs to conduct experiments.

\subsubsection{\textbf{Evaluation}} We use F1-macro and F1-micro to evaluate the performance of HGNNs. We also use the average relative improvement~(ARI) as the metric to evaluate the effectiveness of HDHGR. Given $\mathcal{M}$ the evaluated model set, ARI works as:
\begin{equation}
    \text{ARI}=\frac{1}{|\mathcal{M}|}\sum_{m_i\in\mathcal{M}}\left( \frac{\text{ACC}(m_i(\hat{\mathcal{G}}))- \text{ACC}(m_i(\mathcal{G}))}{\text{ACC}(m_i(\mathcal{G}))}\right)
\end{equation}
\noindent where $\hat{\mathcal{G}}$ is the rewired HG and $\mathcal{G}$ is the original HG. AGR can also be used to evaluate the effectiveness of other HG rewire methods and structure learning methods.

\subsection{Main Performance}

We compare different methods on HGs before and after HDHGR. The results are shown in Table~\ref{tab:main_exp} and Table~\ref{tab:main_exp1}. From the results, we have the following discoveries:

\noindent\textbf{Homophily has a great impact on HGNN performance.} From Table~\ref{tab:main_exp} and Table~\ref{tab:main_exp1}, we find that homophily has a great impact on the performance of HGNNs in the semi-supervised node classification task. HGNNs tend to perform better on homophilous HG, e.g. academic collaborative networks like ACM and DBLP. However, when dealing with non-homophilous HGs like movie-specific social network Actor and IMDB or news network Liar, the performance will degenerate a lot. The reason is that in homophilous HG, the meta-paths tend to have a higher homophily ratio thus the target nodes of the same class are easily aggregated to a similar point in hidden space. While in non-homophilous HG, the complexity measure tends to become very large according to the theory, making the model harder to distinguish different classes. 

\begin{table}[t]
\Large
    \centering 
    \caption{Experimental Results on HGB datasets.}
    \resizebox{.9\linewidth}{!}{
    \begin{tabular}{cccccc}
    \toprule
    \multirow{2}{*}{Model}    & \multirow{2}{*}{/}  & \multicolumn{2}{c}{ACM} & \multicolumn{2}{c}{DBLP} \\
          \cline{3-6}
       \multicolumn{2}{c}{} & Macro-F1 & Micro-F1 & Macro-F1 & Micro-F1  \\
         \midrule
        \multirow{2}{*}{GCN} & origin & $92.17\pm 0.24$ & $92.12\pm 0.23$ & $90.84\pm 0.32$ & $91.47\pm 0.34$  \\
        & HDHGR & $\textbf{92.87}\pm \textbf{0.24}$ & $\textbf{92.80}\pm \textbf{0.25}$ & $\textbf{91.58}\pm \textbf{0.13}$ & $\textbf{92.01}\pm \textbf{0.10}$ \\
        \midrule
         \multirow{2}{*}{GAT} & origin & $92.26\pm 0.94$ & $92.19\pm 0.93$ & $93.83\pm 0.27$ & $93.39\pm 0.30$ \\
         & HDHGR & $\textbf{93.00}\pm \textbf{0.29}$ & $\textbf{92.93}\pm \textbf{0.29}$ & $\textbf{94.31}\pm \textbf{0.12}$ & $\textbf{94.70}\pm \textbf{0.11}$   \\
         \midrule
         \multirow{2}{*}{H2GCN} & origin & $88.79\pm 0.45$ & $88.63\pm 0.48$ & $92.13\pm 0.21$ & $92.34\pm 0.22$ \\
         & HDHGR & $\textbf{90.87}\pm \textbf{0.17}$ & $\textbf{90.76}\pm \textbf{0.18}$ & $\textbf{92.56}\pm \textbf{0.23}$ & $\textbf{92.77}\pm \textbf{0.31}$ \\
         \midrule
         \multirow{2}{*}{LINKX} & origin & $84.75\pm 0.38$ & $84.80\pm 0.38$ & $93.32\pm 0.20$ & $93.24\pm 0.18$\\
         & HDHGR & $ \textbf{89.79}\pm \textbf{0.23}$ & $\textbf{89.71}\pm \textbf{0.25}$ & $\textbf{93.75}\pm \textbf{0.24}$ & $\textbf{93.69}\pm \textbf{0.26}$ \\
         \midrule
         \multirow{2}{*}{RGCN} & origin & $91.55\pm 0.74$ & $91.41\pm 0.75$ & $91.52\pm 0.50$ & $92.07\pm 0.50$\\
         & HDHGR & $\textbf{91.84}\pm \textbf{0.66}$ & $\textbf{91.78}\pm \textbf{0.67}$ & $\textbf{91.88}\pm \textbf{0.76}$ & $\textbf{92.59}\pm \textbf{0.63}$ \\
         \midrule
         \multirow{2}{*}{HAN}   & origin & $90.89\pm 0.43$ & $90.79\pm 0.43$  & $91.67\pm 0.49$ & $92.05\pm 0.62$   \\
         &  HDHGR   & $\textbf{93.88}\pm \textbf{0.20}$ & $\textbf{93.80}\pm \textbf{0.20}$ & $\textbf{92.03}\pm \textbf{0.44}$ & $\textbf{92.57}\pm \textbf{0.56}$   \\
          \midrule
         \multirow{2}{*}{SHGN} & origin & $93.42\pm 0.44$ & $93.35\pm 0.45$ & $94.01\pm 0.24$ & $94.46\pm 0.22$    \\
         & HDHGR & $\textbf{93.75}\pm \textbf{0.22}$ & $\textbf{93.67}\pm \textbf{0.21}$ & $\textbf{94.43}\pm \textbf{0.20}$ & $\textbf{94.73}\pm \textbf{0.16}$ \\
         \midrule
         ARI & / & $1.97\%$ & $1.97\%$ & $0.50\%$ & $0.62\%$ \\
         \bottomrule
    \end{tabular}
    }
    \label{tab:main_exp1}
\end{table}

\noindent\textbf{HDHGR consistently improves HGNN.} In most cases, HGNNs perform better after the HG is processed by HDHGR.  It demonstrates the effectiveness of HDHGR in finding the correct connection between test target nodes of the same class. Besides, we find such performance gain is consistent with different types of HGNNs, i.e., those with semantic level and instance level aggregation and those with only instance level aggregations, showing that HDHGR is a universal algorithm that is not only suitable for some special kind of HGNNs. We also find that compared with GNNs that are originally designed for homogeneous graphs~(e.g. GCN, GAT, $\text{H}_2\text{GCN}$ and LINKX), HDHGR brings more gain on HGNNs that distinguishes different semantics~(e.g. RGCN, HGT, HAN and MAGNN). It indicates that mining useful semantic level information is helpful for node classification, which is consistent with~\cite{zhao2022space4hgnn}. Moreover, although HGNNs do not always perform better than homogeneous GNNs on HGs, after the process of HDHGR, they sometimes outperform their homogeneous counterparts. One more interesting discovery is that HDHGR helps make classification on long-tailed classes, as the ARI on Macro-F1 is always larger than that on Micro-F1, and Macro-F1 weighs each class equally.

\noindent\textbf{HDHGR is more effective on Non-homophilous HG.} Compared with the result in Table~\ref{tab:main_exp1}, HGNNs have a larger performance gain after the processing of HDHGR in Table~\ref{tab:main_exp}. It indicates that HDHGR can effectively improve the local homophily ratio thus making HGNNs more easily classify the target nodes. As homophilous HG already has a high local homophily ratio, continue increasing the homophily ratio will have little help. We also show the result of the homophily ratio of meta-path subgraphs after being rewired by HDHGR in Table~\ref{tab:homorate}. We find that the homophily ratio increased after HDHGR. It indicates that HDHGR learns node similarity accurately.

\begin{table}[t]
\Huge
    \centering
    \caption{Homophily Ratio After HDHGR.}
    \resizebox{.85\linewidth}{!}{
    \begin{tabular}{ccccccc}
    \toprule
    \multirow{2}{*}{Dataset} & \multirow{2}{*}{Meta-path} & \multicolumn{2}{c}{HR} & \multicolumn{2}{c}{MR}\\
    & & w/o HDHGR & w/ HDHGR & w/o HDHGR & w/ HDHGR \\
    \midrule
       \multirow{3}{*}{FB\_American} & PP & 49.91\% & 50.77\% & \multirow{3}{*}{52.14\%} & \multirow{3}{*}{57.21\%} \\
       & PSP & 52.14\% & 57.21\% & \\ 
       & PMP & 51.81\% & 56.07\% & \\
    \midrule
       \multirow{3}{*}{Actor} & SS & 34.18\% & 94.67\% & \multirow{3}{*}{34.18\%} & \multirow{3}{*}{94.67\%} & \\
       & SWS & 27.16\% & 78.89\% & \\
          & SDS & 27.06\% & 77.91\% & \\
    \midrule
    \multirow{2}{*}{Liar} & NSpN & 21.25\% & 58.54\% &  \multirow{2}{*}{21.25\%} & \multirow{2}{*}{66.21\%} \\
    
    & NCN & 18.85\% & 66.21\% & \\
    \midrule
    \multirow{2}{*}{IMDB} & MAM & 61.41\% & 65.37\% & \multirow{2}{*}{61.41\%} & \multirow{2}{*}{67.07\%} \\
    & MDM & 51.60\% & 67.07\% & \\
    \bottomrule
    \end{tabular}
    }
    \label{tab:homorate}
\end{table}

\subsection{Comparison of Graph Structure Learning and Rewiring Methods}

\begin{table}[t]
    \centering
    \Large
    \caption{Comparisons of HGSL and HGR methods.}
    \resizebox{.9\linewidth}{!}{
    \begin{tabular}{ccccc}
    \toprule
      \multirow{2}{*}{Model}  & \multicolumn{2}{c}{Actor} & \multicolumn{2}{c}{Liar}  \\
      \cline{2-5}
        & Macro-F1 & Micro-F1 & Macro-F1 & Micro-F1 \\
        \midrule
        OriMetapath & $42.18\pm 0.46$ & $58.51\pm 0.21$ & $18.56\pm 2.21$ & $23.68\pm 0.84$ \\
         LDS & $56.22\pm 3.24$ & $62.35\pm 2.89$ & $15.65\pm 1.27$ & $19.76\pm 1.74$\\
        IDGL & $59.64\pm 1.71$ & $65.91\pm 2.73$ & $18.07\pm 1.87$ & $21.11\pm 0.86$\\
        HGSL & $58.35\pm 1.23$ & $63.26\pm 2.54$ & $16.74\pm 2.33$ & $20.32\pm 1.41$ \\
        \midrule
        HDHGR & $\textbf{67.21}\pm \textbf{0.94}$ & $\textbf{75.07}\pm \textbf{0.38}$ & $\textbf{21.01}\pm \textbf{1.24}$ & $\textbf{24.08}\pm \textbf{0.37}$ \\
         \bottomrule
    \end{tabular}
    }
    \label{tab:HGSL}
\end{table}

\begin{table*}[t]
\Large
    \centering
    \caption{Ablation Study with SHGN.}
    \resizebox{.85\linewidth}{!}{
    \begin{tabular}{ccc|cc|cc|cc}
    \toprule
   \multirow{2}{*}{Variants} & \multicolumn{2}{c}{FB\_American} & \multicolumn{2}{c}{Actor} & \multicolumn{2}{c}{Liar} & \multicolumn{2}{c}{IMDB} \\
   & Macro-F1 & Micro-F1 & Macro-F1 & Micro-F1 & Macro-F1 & Micro-F1 & Macro-F1 & Micro-F1 \\
    \midrule
       HDHGR w/o Label  & $72.97\pm 1.66$ & $75.44\pm 1.67$ & $72.05\pm 0.49$ & $80.37\pm 0.30$ & $19.66\pm 1.08$ & $23.42\pm 0.58$ & $52.33\pm 0.96$ & $53.44\pm 0.75$  \\
       HDHGR w/o Attr  & $75.02\pm 1.20$ & $77.17\pm 1.91$ & $71.43\pm 0.60$ & $79.53\pm 0.24$ & $18.35\pm 0.55$ & $23.26\pm 0.38$ & $54.65\pm 0.67$ & $55.68\pm 0.72$ \\
       HDHGR w/o MO  & $74.04\pm 0.48$ & $76.46\pm 0.25$ & $71.90\pm 0.42$ & $79.70\pm 0.21$ & $19.40\pm 0.35$ & $23.53\pm 0.37$ & $54.51\pm 0.37$ & $55.77\pm 0.54$  \\
       \midrule
       HDHGR  & $\textbf{76.03}\pm \textbf{1.65}$ & $\textbf{77.84}\pm \textbf{1.50}$ & $\textbf{72.73}\pm \textbf{0.35}$ & $\textbf{81.36}\pm \textbf{0.37}$ & $\textbf{22.29}\pm \textbf{0.72}$ & $\textbf{23.82}\pm \textbf{1.08}$ & $\textbf{55.69}\pm \textbf{0.61}$ & $\textbf{56.56}\pm \textbf{0.48}$  \\
       \bottomrule
    \end{tabular}
    }
    \label{tab:ablation}
\end{table*}

We compare the performance of graph structure learning methods LDS, IDGL, and HGSL with HDHGR. Nevertheless, we also compare a baseline that uses the original meta-paths denoted by OriMetapath. The results are shown in Table~\ref{tab:HGSL}. We use GCN as the base model. We find that directly adding original meta-path instances leads to poor performance compared to the result by HDHGR. A proper explanation is that the original meta-path contains lots of instances connecting nodes from different classes, and HDHGR can prune redundant meta-path instances and only remain instances that are useful for node classification. Besides, the heterogeneous graph structure learning method HGSL shows comparative performance compared with homogeneous graph structure learning methods. Moreover, we also find that HDHGR outperforms graph structure learning methods LDS, IDGL, and HGSL, indicating its effectiveness in alternating HG structure to facilitate HGNN learning.  

\subsection{Ablation Study}

We conduct an ablation study to show the influence of different modules, i.e., the label information, the attribute information, and the multi-objective optimization of HDHGR. For HDHGR w/o Label, we only use the attribute similarity matrix to learn MSL. For HDHGR w/o Attr, we only use the label similarity matrix to learn MSL. As for HDHGR w/o MO, we uniformly assign weights to the loss for each meta-path. The result is shown in Table~\ref{tab:ablation}. We find that label information, attribute information, and multi-objective optimization all have positive effects on HDHGR. Among the four datasets, FB\_American and IMDB are more influenced by the label information, while Actor and Liar are more influenced by the attribute information. Although the two kinds of similarities are all effective, we also need multi-objective optimization to weigh different meta-paths to perform better. 

\subsection{Hyper-parameter Study}

\subsubsection{\textbf{Influence of the Degree Threshold $\alpha$}} 
\begin{figure}[t]
    \centering
    \subfigure[FB-American]{
    \includegraphics[width=.41\linewidth]{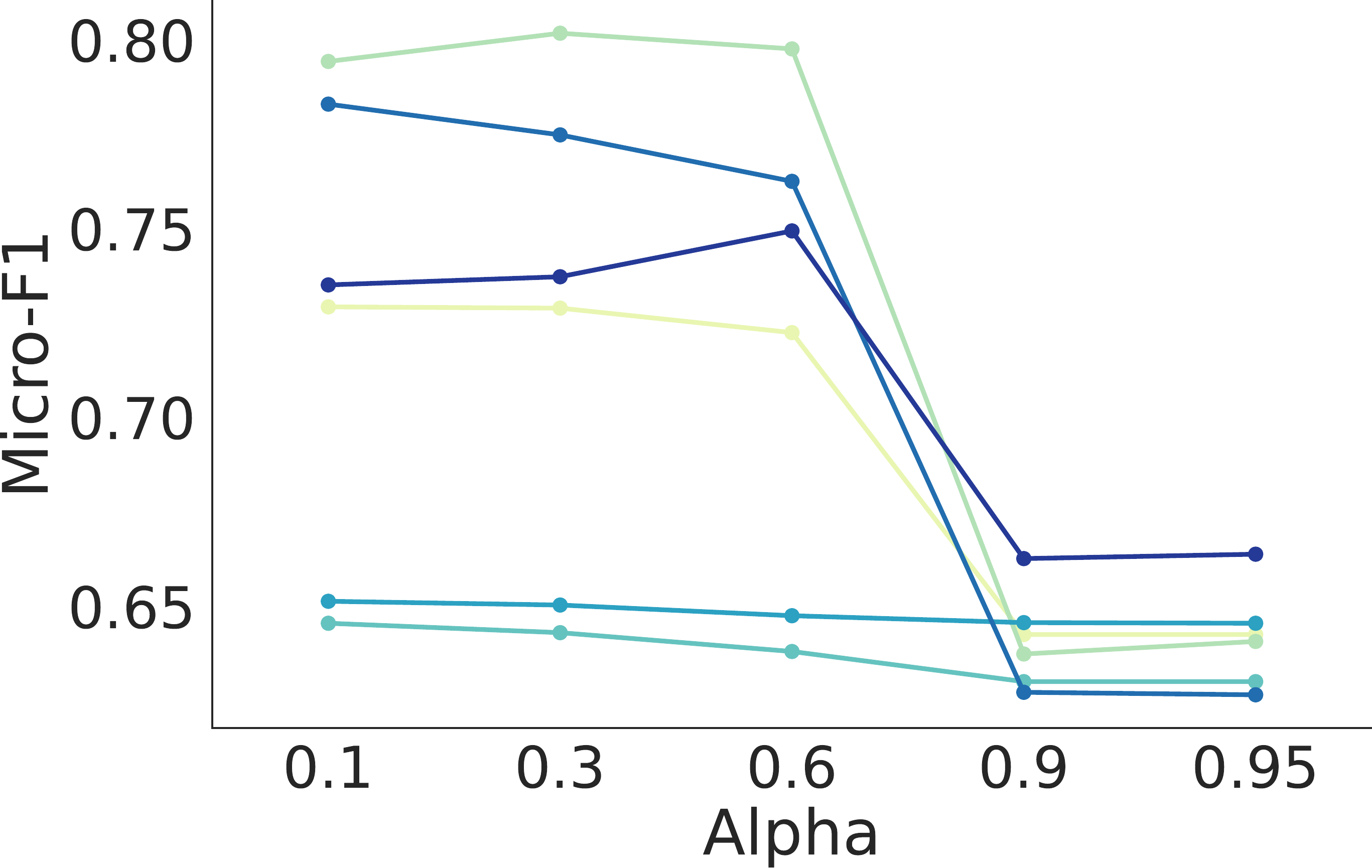}
    }
    \subfigure[Actor]{
    \includegraphics[width=.54\linewidth]{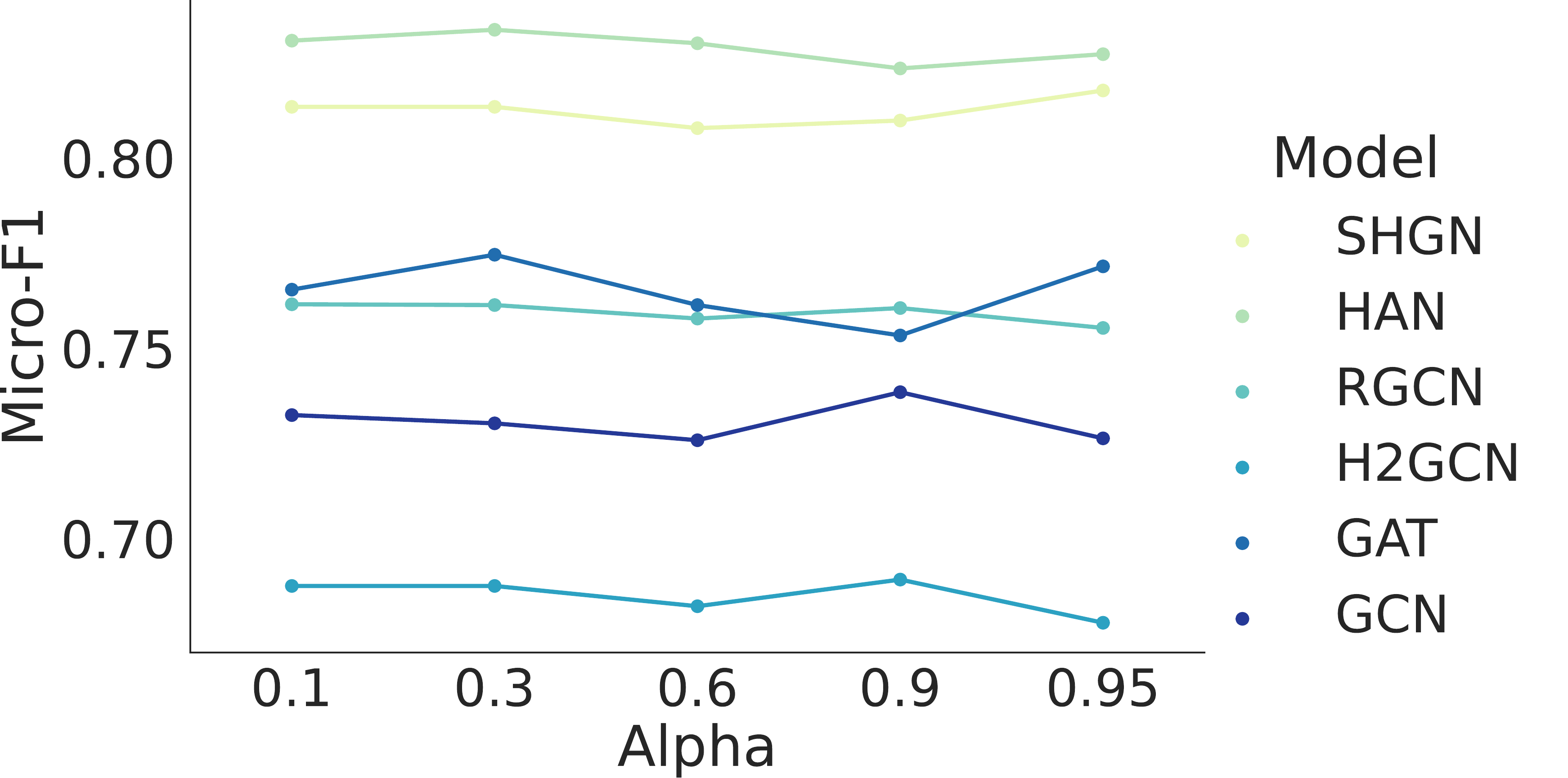}
    }
    \caption{Hyper-parameter Study of $\alpha$.}
    \label{fig:hyper_alpha}
\end{figure}

From Figure~\ref{fig:hyper_alpha}, we find that the minimum degree thresholds $\alpha$ influences the performance of HGNNs. Besides, such influence varies according to different HGNNs and datasets. In most cases, the middle value of 0.3 to 0.6 generates better results. It shows that $\alpha$ should not be too small or too large otherwise it may degenerate the performance.

\subsubsection{\textbf{Influence of the Growing Threshold $K$ and Minimum Similarity $\epsilon$}} $K$ and $\epsilon$ influences the density of the rewired subgraph. From the experimental result, We observe that the performance usually increases when increasing $\epsilon$ with fixed K, while decreases when increasing $K$ with fixed $\epsilon$, demonstrating the effectiveness and robustness of the rewired HG learned by HDHGR.

\begin{figure}[t]
    \centering
    \subfigure[FB-American]{
    \includegraphics[width=.475\linewidth]{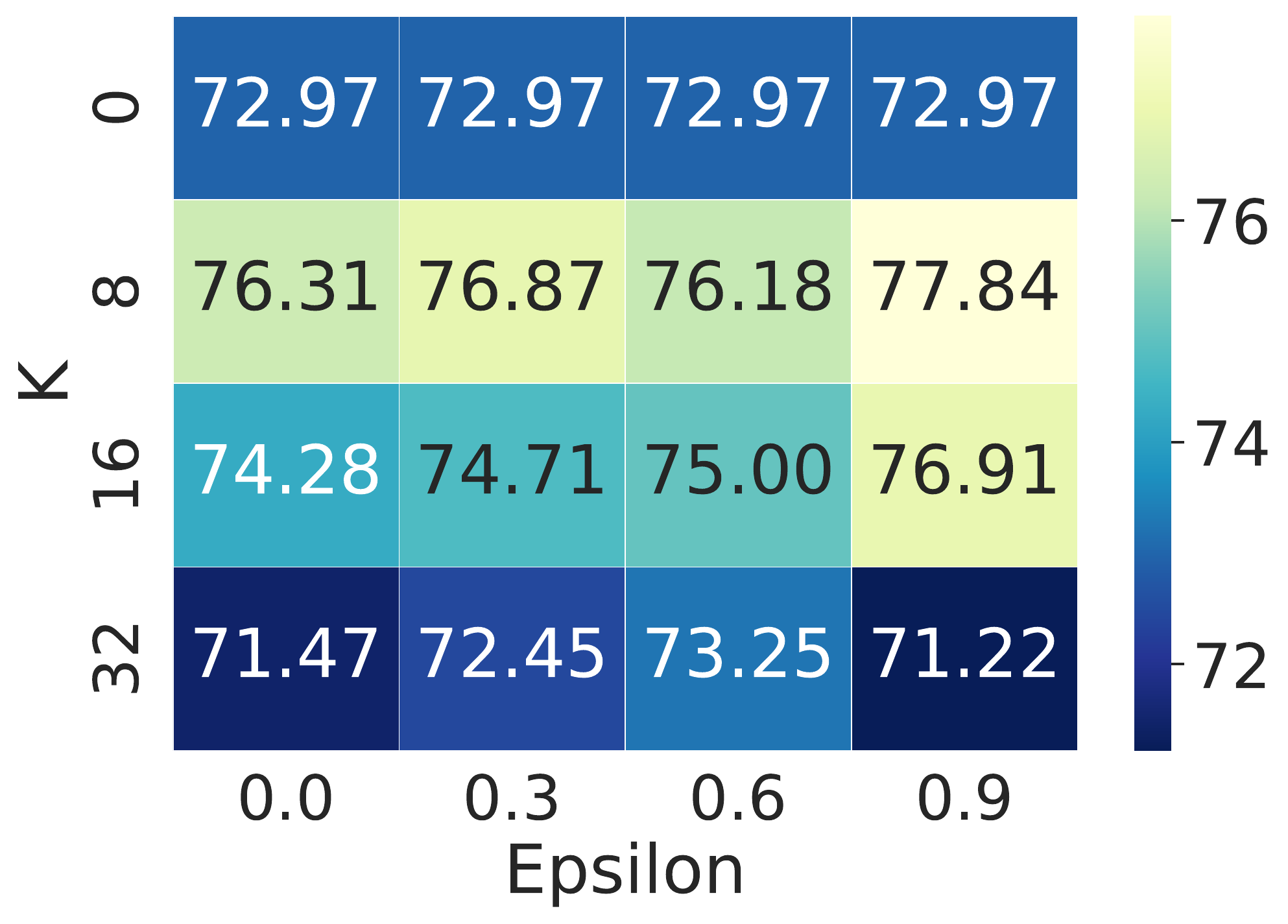}
    }
    \subfigure[Actor]{
    \includegraphics[width=.475\linewidth]{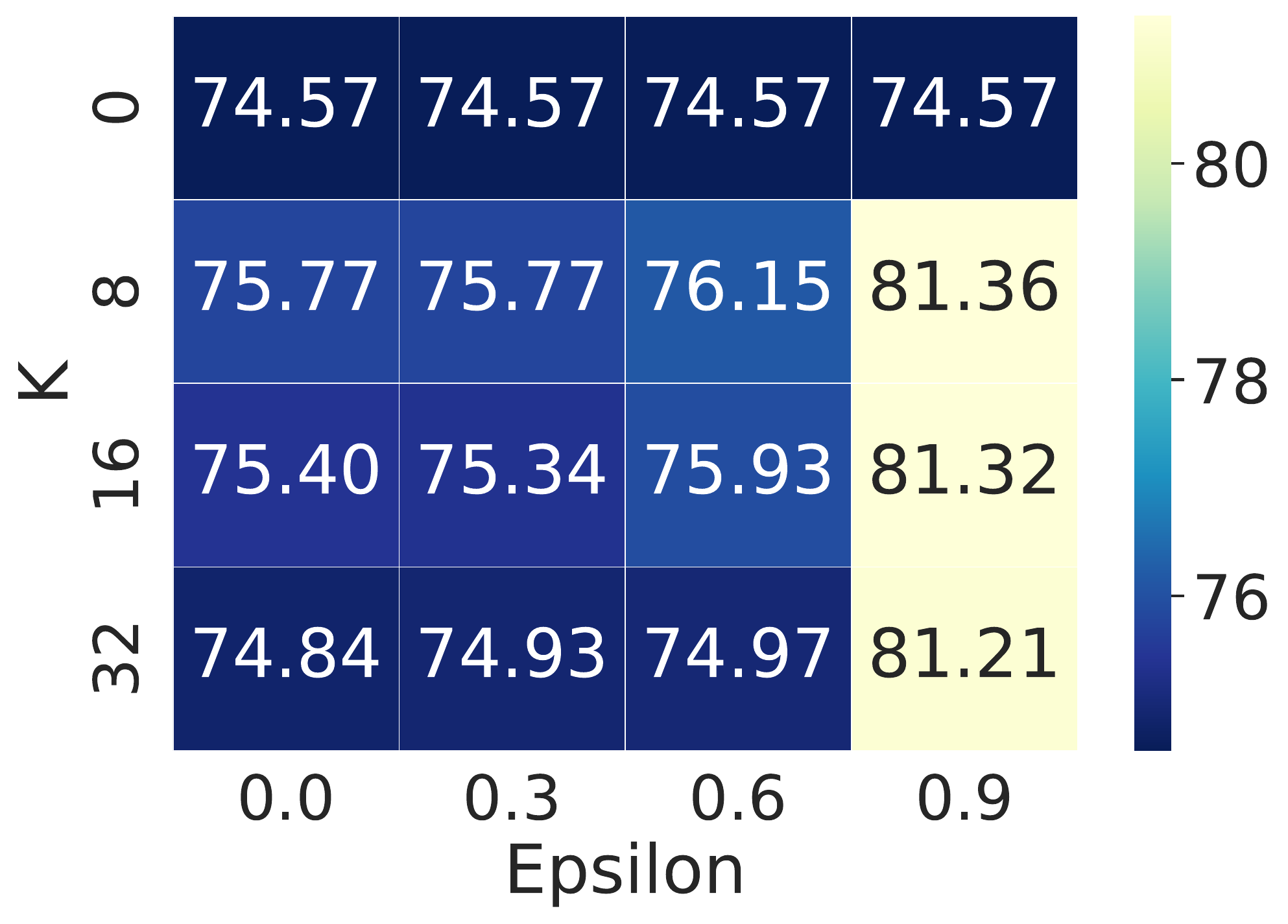}
    }
    \caption{Hyper-parameter Study of $K$ and $\epsilon$.}
    \label{fig:hyper_k_e}
\end{figure}


\section{Related Work}

\noindent\textbf{Heterogeneous Graph Neural Networks.} There is a broad range of works on HGNNs in recent years~\cite{wang2019heterogeneous,fu2020magnn,lv2021we,yang2022simple,zhao2022space4hgnn,zhang2019heterogeneous,yu2022multiplex,hu2020heterogeneous,yun2019graph,zhu2019relation,hong2020attention,jin2021heterogeneous}. Many HGNNs are designed to enhance the GNN architecture with the capability of capturing the node and edge heterogeneous contextual signals. For example, HetGNN~\cite{zhang2019heterogeneous} jointly encodes the graph
topology and context heterogeneity for representation learning. HAN~\cite{wang2019heterogeneous} and MAGNN~\cite{fu2020magnn} design message passing along meta-path to integrate semantic patterns. GTN~\cite{yun2019graph} designs a learnable gate to assign weights to different meta-paths to automatically select useful meta-paths. Recently, some works show that directly applying message passing with type regularized information on HG without predefined meta-paths achieves better performance~\cite{hong2020attention,lv2021we, bi2022company,yu2022multiplex}. Space4HGNN~\cite{zhao2022space4hgnn} studies the design space for HGNN. Although these researches directly focus on the model design, there is a lack of a study of what kind of heterogeneous graphs the HGNNs are good at handling.
\newline

\noindent\textbf{Heterophily and Graph Neural Network.} Recent works show that GNNs degenerate performance on non-homophilous graphs~\cite{zhu2020beyond,zhu2021graph,yan2021two,wang2022powerful,zheng2022graph,ma2022meta,du2022gbk,zhu2022does,jin2021universal,park2022deformable,li2022graph,yang2022graph,yang2021diverse,fang2022polarized}. This
issue is also known in classical semi-supervised learning~\cite{peel2017graph}. To address this issue, several GNN designs for handling heterophilous connections have been proposed~\cite{abu2019mixhop,bo2021beyond,dong2021adagnn,li2021beyond, bi2022mm,Pei2020GeomGCNGG}. GGCN~\cite{yan2021two} recently discusses the connection between heterophily and over-smoothing for GNNs, and designs message-passing strategies to address both issues. ~\cite{loveland2022graph} studies how locally occurring heterophily affects the fairness of GNNs. Zhu et al.~\cite{zhu2022does} studies the formal connection between heterophily and the robustness of GNNs. Here we focus on a simple yet powerful design that significantly improves performance under heterophily~\cite{zhu2020beyond} and can be readily incorporated into HGNNs.
\newline

\noindent\textbf{Graph Rewiring and Structure Learning.} Recent graph structure learning methods~\cite{franceschi2019learning,jin2020graph,zhu2021deep,zhao2021heterogeneous,liu2022towards,sun2022graph,chen2020iterative,peng2022reverse} aims to learn a graph structure from the original graph or noisy data points that reflect data relationships. Different from GSL, which learns a graph structure via training GNN, graph rewiring~(GR) aims to decouple the structure learning and the downstream task. It adjusts the graph structure first to further facilitate downstream model training and inference~\cite{topping2021understanding,kenlay2021stability,bi2022make}. A recent study uses rewiring to reduce the bottleneck of a graph or make adversarial attacks.

\section{Conclusion}

In this paper, we study how the homophily property of HGs affects HGNN performance. We find that HGNN tends to perform better on homophilous HGs. To help HGNN better process non-homophilous HG, we propose HDHGR, a homophily-oriented deep heterogeneous graph rewiring method to improve the meta-paths subgraph homophily ratio thus making HGNNs generalize better. Extensive experiments demonstrate the effectiveness of our method.  In the future, we will consider extending the method to dynamic heterogeneous graphs. Besides, we will consider developing rewiring methods for heterogeneous hypergraphs. Moreover, we will consider rewiring the graph for tasks that beyond node classification. For example, link prediction and graph classification.

\newpage

\bibliographystyle{ACM-Reference-Format}
\balance{
\bibliography{main}}

\newpage

\nobalance
\appendix

\section{Theoretical Analysis}
\label{app:theory}


\noindent\textbf{Theorem 1.} \textit{On the heterogeneous graph $\mathcal{G}=(\mathcal{V},\mathcal{E},\phi,\psi)$ with metapath set $\mathcal{R}$, we consider a binary classification task under the condition that the neighbors of each node $v_i\in \mathcal{V}$ have the probability of $P^r$ to belong to the same class with $v_i$ under metapath $r$. 
\textbf{For an arbitrary metapath type $r\in\mathcal{R}$, when $P^r\rightarrow 1$, the  upper bound of the HGNN's generalization ability reach optimal.}
 }
 \begin{proof}
 We prove the theorem by analyzing the \textbf{exact lower bound} of the complexity measure $\mathcal{C}$ of the general heterogeneous graph convolutional layer. Based on the Consistency of Representations~\cite{natekar2020representation} and  Fisher discriminant analysis \cite{fisher1936use}, which all use the ratio of the within-class variance to the between-class variance as an indicator, we rewrite the Consistency of Representations for the convenience of theoretical analysis:
 \begin{equation}
\small
     \mathcal{C} = \frac{1}{k}\sum_{i=0}^{k-1}\mathop{\text{max}}_{i\neq j}\frac{S_i^2+S_j^2}{M_{i,j}^2}
 \end{equation}
    Firstly we derive $\mu_{\mathcal{O}_{0}}$ and $\mu_{\mathcal{O}_{1}}$
    \begin{equation}
    \small
        \begin{split}
            \mu_{\mathcal{O}_0}&=\mathbb{E}\left (\sum_{r\in\mathcal{R}} \textbf{W}_{r}\sum_{j\in\mathcal{N}_{r}(v_i)}\frac{1}{|\mathcal{N}_{r}(v_i)|} \textbf{X}^{(j)} \right) \\
            &=\sum_{r\in\mathcal{R}}\mathbb{E}\left ( \textbf{W}_{r}\sum_{j\in\mathcal{N}_{r}(v_i)}\frac{1}{|\mathcal{N}_{r}(v_i)|} X^{(j)} \right) \\
            &=\sum_{r\in\mathcal{R}} \textbf{W}_{r}\left(P^{r}\mu_{\textbf{X}_0}+(1-P^{r})\mu_{\textbf{X}_1}\right) \\
        \end{split}
    \end{equation}
    Similarly, we have:
    \begin{equation}
    \small
        \begin{split}
            \mu_{\mathcal{O}_{1}}&=\mathbb{E}\left (\sum_{r\in\mathcal{R}} \textbf{W}_{r}\sum_{j\in\mathcal{N}_{r}(v_i)}\frac{1}{|\mathcal{N}_{r}(v_i)|} X^{(j)} \right) \\
            &=\sum_{r\in\mathcal{R}}\mathbb{E}\left ( \textbf{W}_{r}\sum_{j\in\mathcal{N}_{r}(v_i)}\frac{1}{|\mathcal{N}_{r}(v_i)|} X^{(j)} \right) \\
            &=\sum_{r\in\mathcal{R}} \textbf{W}_{r}\left(P^{r}\mu_{\textbf{X}_1}+(1-P^{r})\mu_{\textbf{X}_0}\right) \\
        \end{split}
    \end{equation}
\noindent Then we can derive $M_{0,1}$:
\begin{equation}
\label{eq:M_01}
\small
    \begin{split}
        M_{0,1}^2=&\left\vert\left\vert\mu_{\mathcal{O}_{0}}-\mu_{\mathcal{O}_{1}}\right|\right|^2\\
        =&\left|\left| \sum_{r\in\mathcal{R}} \textbf{W}_{r}\left(P^{r}\mu_{\textbf{X}_0}+(1-P^{r})\mu_{\textbf{X}_1}\right )-\sum_{r\in\mathcal{R}} \textbf{W}_{r}\left(P^{r}\mu_{\textbf{X}_1}+(1-P^{r})\mu_{\textbf{X}_0}\right ) \right|\right|^2 \\
        =& \left|\left|\sum_{r\in\mathcal{R}}\left ( 2P^{r}-1 \right ) \textbf{W}_{r}\left(\mu_{\textbf{X}_0}-\mu_{\textbf{X}_1}\right) \right|\right|^2 \\
        \le & \sum_{r\in\mathcal{R}}\left(2P^{r} -1 \right)^2 \cdot\left\vert\left\vert\textbf{W}_{r}\left(\mu_{\textbf{X}_0}-\mu_{\textbf{X}_1}\right) \right\vert\right\vert^2 \\
    \end{split}
\end{equation}
\noindent The final result is derived by Jensen's Inequality.  



Then we can calculate the intra-class variance $S_0^2$ for class $0$.
\begin{equation}
\footnotesize
\begin{aligned}
    S_0^2 &= \mathbb{E}\left(<\mathcal{O}_0^{(i)} - \mu_{\mathcal{O}_0}, \mathcal{O}_0^{(i)} - \mu_{\mathcal{O}_0}> \right)\\
    &=\mathbb{E}\left( \sum_{r_1, r_2 \in \mathcal{R}} P^{r_1}P^{r_2}\cdot(X_0-\mu_{X_0})^T\cdot \mathbf{W}_{r_1}^T\mathbf{W}_{r_2} )\cdot (X_0-\mu_{X_0}) \right)\\
    &\quad+ \mathbb{E}\left( \sum_{r_1, r_2 \in \mathcal{R}} (1-P^{r_1})(1-P^{r_2})\cdot(X_0-\mu_{X_0})^T\cdot \mathbf{W}_{r_1}^T\mathbf{W}_{r_2} )\cdot (X_0-\mu_{X_0}) \right)\\
    &=\mathbb{E}\left((X_0-\mu_{X_0})^T\cdot \big(\sum_{r\in\mathcal{R}} P^{r} \mathbf{W}_{r}\big)^T\cdot\big(\sum_{r\in \mathcal{R}} P^{r}\mathbf{W}_{r}\big) \cdot (X_0-\mu_{X_0}) \right) \\
    &\quad+\mathbb{E}\left((X_1-\mu_{X_1})^T\cdot \big(\sum_{r\in\mathcal{R}} (1-P^{r}) \mathbf{W}_{r}\big)^T\cdot\big(\sum_{r\in \mathcal{R}} (1-P^{r})\mathbf{W}_{r}\big) \cdot (X_1-\mu_{X_1}) \right) \\
\end{aligned}
\end{equation}
Then we denote $X_0-\mu_{X_0}$ as $\Delta X_0$ and $X_1-\mu_{X_1}$ as $\Delta X_1$. Then we denote $\sum_{r\in \mathcal{R}} P^{r}\mathbf{W}_{r}$ as matrix $A$ and $\sum_{r\in \mathcal{R}} (1-P^{r})\mathbf{W}_{r}$ as matrix $B$. And we can further simplify the equation:
\begin{equation}
    \footnotesize
    S_0^2 =\mathbb{E}\left( \sum_{r_1, r_2 \in \mathcal{R}} (\Delta X_0)^T\cdot A^T A \cdot \Delta X_0 \right) + \mathbb{E}\left( \sum_{r_1, r_2 \in \mathcal{R}} (\Delta X_1)^T\cdot B^T B \cdot \Delta X_1 \right)
\end{equation}
Similarity, $S_1^2$ is:
\begin{equation}
    \footnotesize
    S_1^2 =\mathbb{E}\left( \sum_{r_1, r_2 \in \mathcal{R}} (\Delta X_0)^T\cdot B^T B \cdot \Delta X_0 \right) + \mathbb{E}\left( \sum_{r_1, r_2 \in \mathcal{R}} (\Delta X_1)^T\cdot A^T A \cdot \Delta X_1 \right)
\end{equation}
We can further simplify the above equations with the inequality $x^T\cdot(A^TA+B^TB)\cdot x\geq x^T\cdot\left(\frac{1}{2}\cdot(A+B)^T(A+B)\right)\cdot x$, which is rigorous and takes the equal sign when $A$ and $B$ are equal (i.e., $\sum_{r\in \mathcal{R}} P^{r}\mathbf{W}_{r}=\sum_{r\in \mathcal{R}} (1-P^{r})\mathbf{W}_{r}$.
Besides, we have $A+B=\sum_{r\in\mathcal{R}} \mathbf{W}_{r}$, and the result of $S_0^2+S_1^2$ can be written as follows:
\begin{equation}
\label{eq:intra_var_lower_bound}
\footnotesize
    \begin{aligned}
        S_0^2+S_1^2 &= \mathbb{E}\left[\Delta X_0^T\cdot \Big( A^TA + B^TB  \Big)  \cdot \Delta X_0\right]\\
        &\qquad+ \mathbb{E}\left[\Delta X_1^T\cdot \Big( B^TB +A^TA  \Big)  \cdot \Delta X_1\right]\\
        &\geq \frac{1}{2}\mathbb{E}\left[\Delta X_0^T\cdot \Big( A + B  \Big)^T\Big( A + B  \Big)  \cdot \Delta X_0\right] \\
        &\qquad + \frac{1}{2}\mathbb{E}\left[\Delta X_1^T\cdot \Big( A+B  \Big)^T\Big( A + B  \Big)  \cdot \Delta X_1\right]\\
        &\geq \frac{1}{2}\mathbb{E}\left[\Delta X_0^T\cdot  \Big(\sum_{r\in\mathcal{R}} \mathbf{W}_{r}\Big)^T\Big(\sum_{r\in\mathcal{R}} \mathbf{W}_{r}\Big)  \cdot \Delta X_0\right]\\
        &\qquad + \frac{1}{2}\mathbb{E}\left[\Delta X_1^T\cdot  \Big(\sum_{r\in\mathcal{R}} \mathbf{W}_{r}\Big)^T\Big(\sum_{r\in\mathcal{R}} \mathbf{W}_{r}\Big)  \cdot \Delta X_1\right]
    \end{aligned}
\end{equation}
For simplification, we further define a mathematical operator, denoted as $\Gamma(\cdot)$:
\begin{equation}
\small
    \nonumber
    \Gamma(X) = X^TX
\end{equation}
Then we further simplify Eq.~\ref{eq:intra_var_lower_bound} as follows:
\begin{equation}
\small
     S_0^2+S_1^2 \geq \frac{1}{2}\mathbb{E}\left[\Delta X_0^T\cdot  \Gamma\left(\sum_{r\in\mathcal{R}} \mathbf{W}_{r}\right)  \cdot \Delta X_0\right]
     + \frac{1}{2}\mathbb{E}\left[\Delta X_1^T\cdot  \Gamma\left(\sum_{r\in\mathcal{R}} \mathbf{W}_{r}\right)  \cdot \Delta X_1\right]
\end{equation}
And we get the exact  lower bound $\mathcal{C}^{lower}$ of the complexity measure $\mathcal{C}$:
\begin{equation}
\footnotesize
    \begin{aligned}
        \mathcal{C} &= \frac{S_0^2+S_1^2}{M_{0,1}^2}\\
        &\geq \frac{\frac{1}{2}\mathbb{E}\left[\Delta X_0^T\cdot  \Gamma\left(\sum_{r\in\mathcal{R}} \mathbf{W}_{r}\right)  \cdot \Delta X_0\right] + \frac{1}{2} \mathbb{E}\left[\Delta X_1^T\cdot  \Gamma\left(\sum_{r\in\mathcal{R}} \mathbf{W}_{r}\right)  \cdot \Delta X_1\right]}{\sum_{r\in\mathcal{R}}\left(2P^{r} -1 \right)^2 \cdot\left\vert\left\vert\textbf{W}_{r}\left(\mu_{\textbf{X}_0}-\mu_{\textbf{X}_1}\right) \right\vert\right\vert^2}
    \end{aligned}
\end{equation}
Then the derivative of the exact lower bound $\mathcal{C}^{lower}$ according to $P^{r^\prime}, r^\prime\in \mathcal{R}$ is:
\begin{equation}
    \label{eq:derivative}
    \footnotesize
    \begin{aligned}
        \frac{\partial\mathcal{C}^{lower}}{\partial{P^{r^\prime}}} &=(2P^{r^\prime}-1)\cdot\frac{4 S_{lower}^2\cdot \left\vert\left\vert\textbf{W}_{r}\left(\mu_{\textbf{X}_0}-\mu_{\textbf{X}_1}\right) \right\vert\right\vert^2 }{\left(\sum_{r\in\mathcal{R}}\left( 1-2P^{r} \right)^2\big\vert\big\vert\textbf{W}_{r}\left(\mu_{\textbf{X}_0}-\mu_{\textbf{X}_1}\right) \big\vert\big\vert^2\right)^2}\\
        &=(2p^{r^\prime}-1)\cdot\mathcal{O}_{+}
    \end{aligned}
\end{equation}
where 
\begin{equation}
\footnotesize
    \mathcal{O}_+ = \frac{4 S_{lower}^2\cdot \left\vert\left\vert\textbf{W}_{r}\left(\mu_{\textbf{X}_0}-\mu_{\textbf{X}_1}\right) \right\vert\right\vert^2 }{\left(\sum_{r\in\mathcal{R}}\left( 1-2P^{r} \right)^2\big\vert\big\vert\textbf{W}_{r}\left(\mu_{\textbf{X}_0}-\mu_{\textbf{X}_1}\right) \big\vert\big\vert^2\right)^2} > 0
\end{equation}
and
\begin{equation}
    \footnotesize
    S_{lower}^2=\frac{1}{2}\mathbb{E}\left[\Delta X_0^T\cdot  \Gamma\left(\sum_{r\in\mathcal{R}} \mathbf{W}_{r}\right)  \cdot \Delta X_0\right] + \frac{1}{2} \mathbb{E}\left[\Delta X_1^T\cdot  \Gamma\left(\sum_{r\in\mathcal{R}} \mathbf{W}_{r}\right)  \cdot \Delta X_1\right]
\end{equation}
, which is constant to $P^r$. 
According to the derivative in Eq.~\ref{eq:derivative}, it is obvious that the exact lower bound of complexity measure $\mathcal{C}^{lower}$ reaches the minimum value when $P^r=1$. Besides, it is continuous and decreases monotonically around $P^r=1$.
Based on the proof, we conclude that: \textbf{Given an arbitrary metapath type $r\in\mathcal{R}$, when $P^r\rightarrow 1$, the  upper-bound of the HGNN's generalization ability reach optimal.} This motivates us to reduce the heterophily degree of heterogeneous graph with graph rewiring.

\end{proof}

\section{Detailed Statistics of the Datasets}

The meta-paths statistics are shown in Table~\ref{tab:statistics_detail}.

\begin{table}[t]
    \centering
    \caption{Meta-path Statistics.}
    \resizebox{.9\linewidth}{!}{
    \begin{tabular}{cccc}
    \toprule
      Dataset &  Meta-path & HR~(\%) & Meaning  \\
      \midrule
        FB-American & \makecell[c]{PP \\ PSP \\ PMP} & \makecell[c]{49.91 \\ 52.14 \\ 51.81} & \makecell[c]{P: person \\ S: status \\ M: major \\ H: house \\ } \\
        \midrule
        Actor & \makecell[c]{SS \\ SWS \\ SDS} & \makecell[c]{34.18 \\ 27.16 \\ 27.06} & \makecell[c]{S: star \\ W:writer \\ D:director}  \\
        \midrule
        Liar & \makecell[c]{NSpN \\ NSuN \\ NCN}  & \makecell[c]{21.25 \\ 18.14 \\ 18.85} & \makecell[c]{N: news \\ Sp: speaker \\ Su: subject \\ C: context} \\
        \midrule
        IMDB & \makecell[c]{MDM \\ MAM} & \makecell[c]{61.41 \\ 44.43} & \makecell[c]{M: movie \\ D: director \\ A: actor} \\
        \midrule
        ACM & \makecell[c]{PcP, PrP \\ PAP \\ PSP \\ PTP} & \makecell[c]{87.95 \\ 79.36 \\ 63.98 \\ 33.36} & \makecell[c]{P: paper \\ A: author \\ S: subject \\ T: term \\ c: citation relation \\ r: reference relation}\\
        \midrule
        DBLP & \makecell[c]{APC \\ APTPA \\ APVPA} & \makecell[c]{79.88 \\ 32.45  \\ 66.97} & \makecell[c]{A: author \\ P: paper \\ T: term \\ V: venue} \\
    \bottomrule
    \end{tabular}}
    \label{tab:statistics_detail}
\end{table}

\section{Extensive Experimental Reulsts}

\subsection{Hyper-parameter Study on $\alpha$}

\begin{figure}[t]
    \centering
    \subfigure[Liar]{
    \includegraphics[width=.41\linewidth]{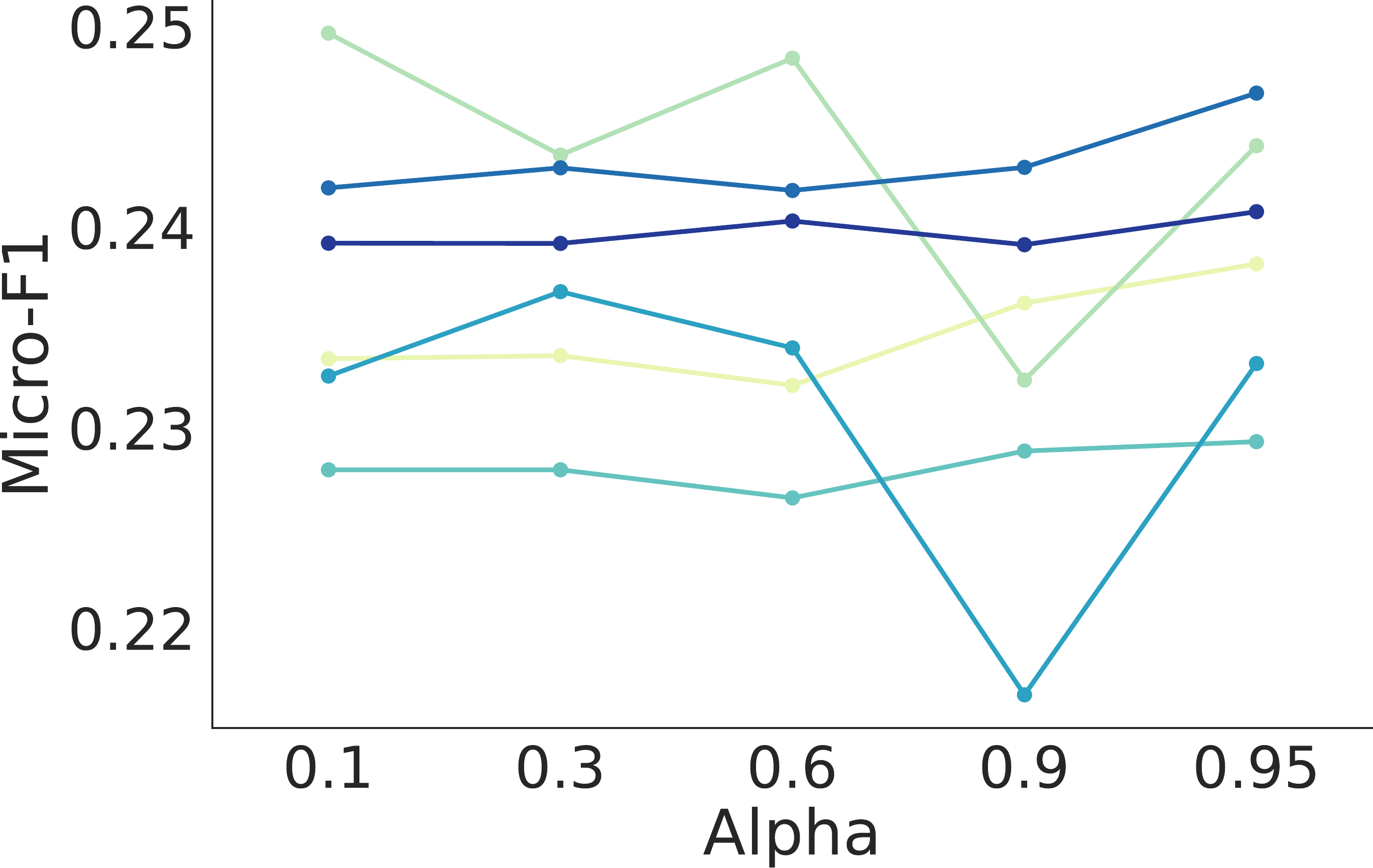}
    }
    \subfigure[IMDB]{
    \includegraphics[width=.54\linewidth]{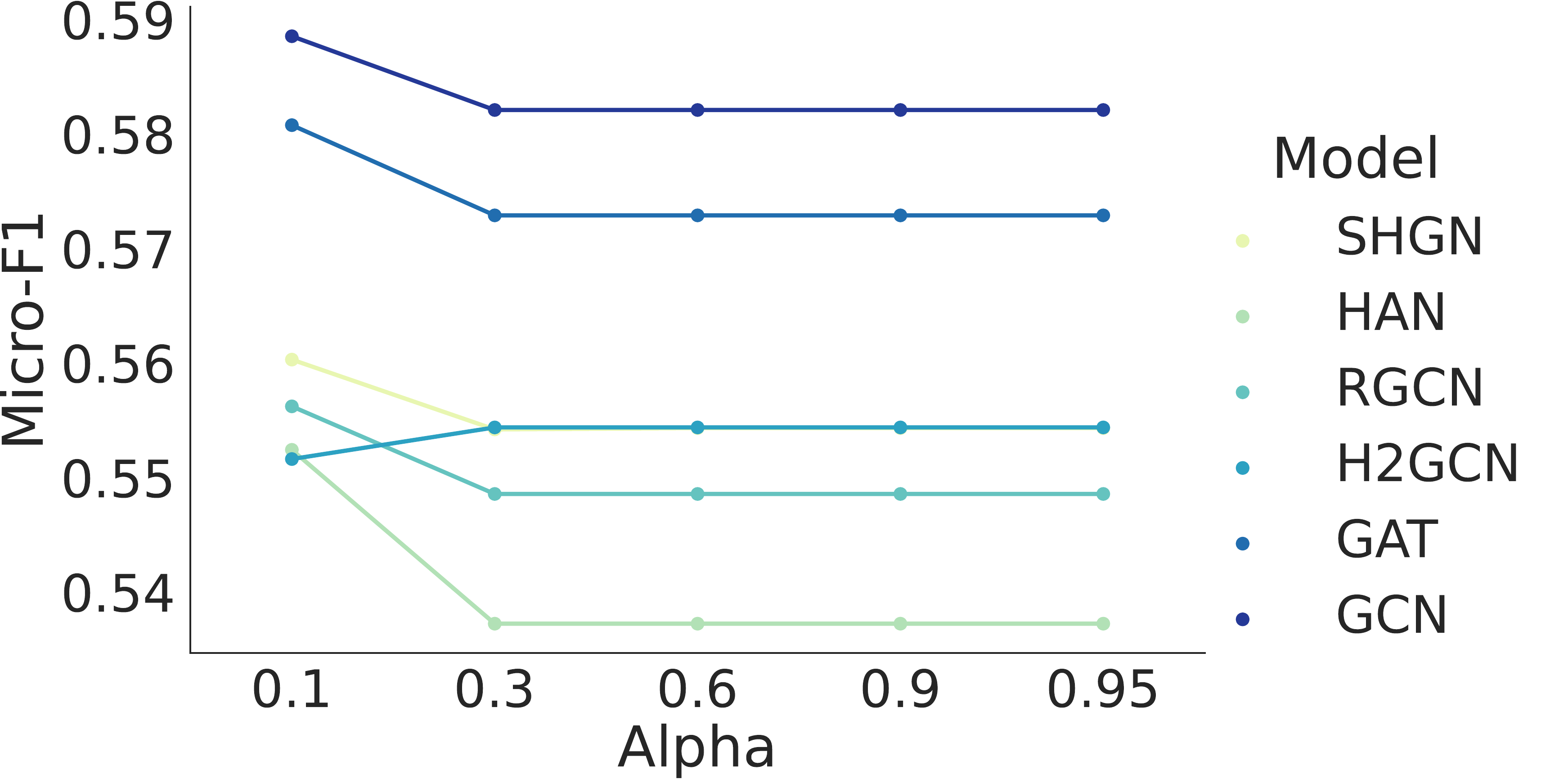}
    }
    \caption{Hyper-parameter Study of $\alpha$.}
    \label{fig:hyper_alpha1}
\end{figure}

The results on dataset Liar and IMDB are shown in Figure~\ref{fig:hyper_alpha1}. We can have similar conclusions in the former discussions.

\subsection{Hyper-parameter Study on $K$ and $\epsilon$}

\begin{figure}[t]
    \centering
    \subfigure[Liar]{
    \includegraphics[width=.475\linewidth]{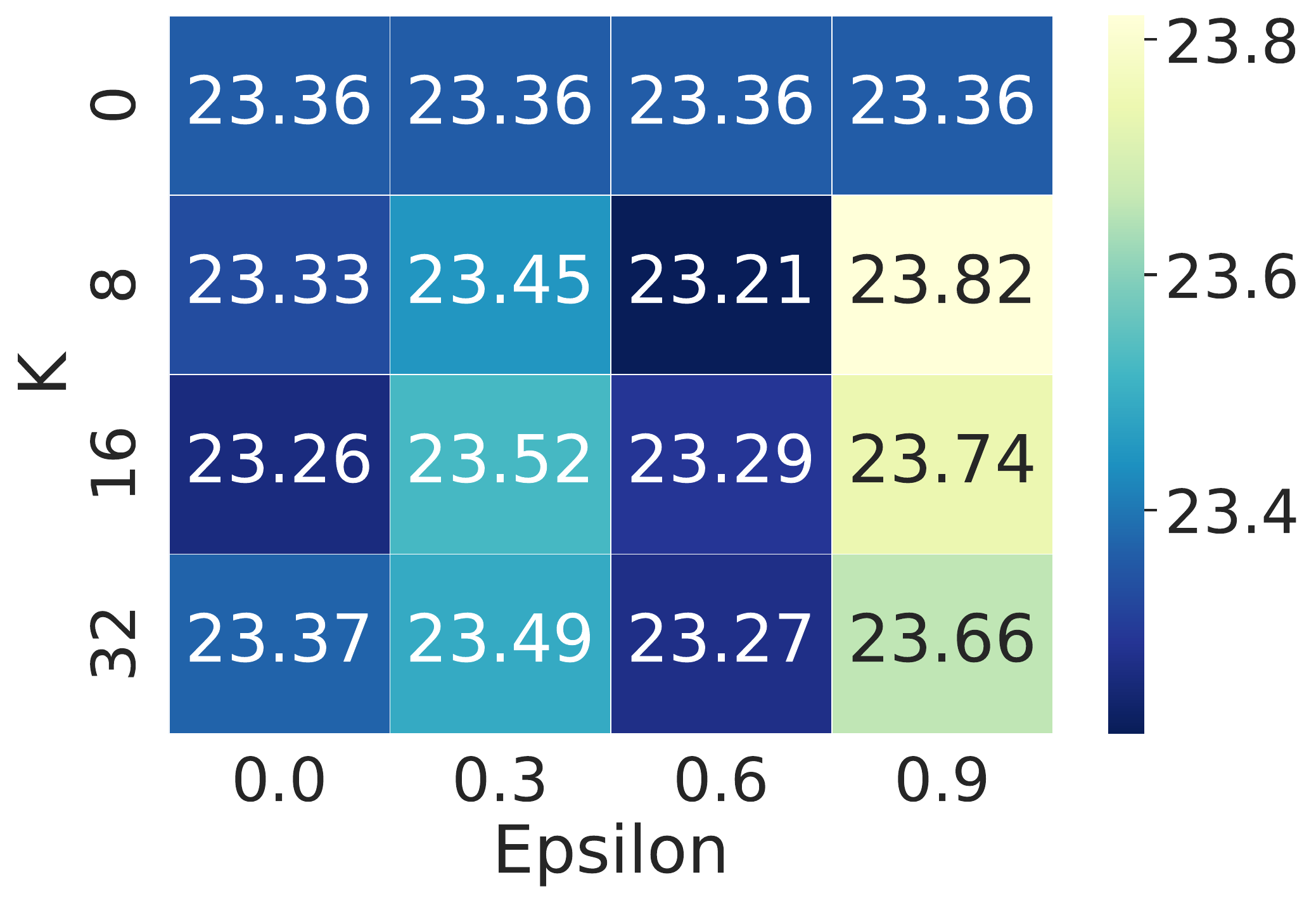}
    }
    \subfigure[IMDB]{
    \includegraphics[width=.475\linewidth]{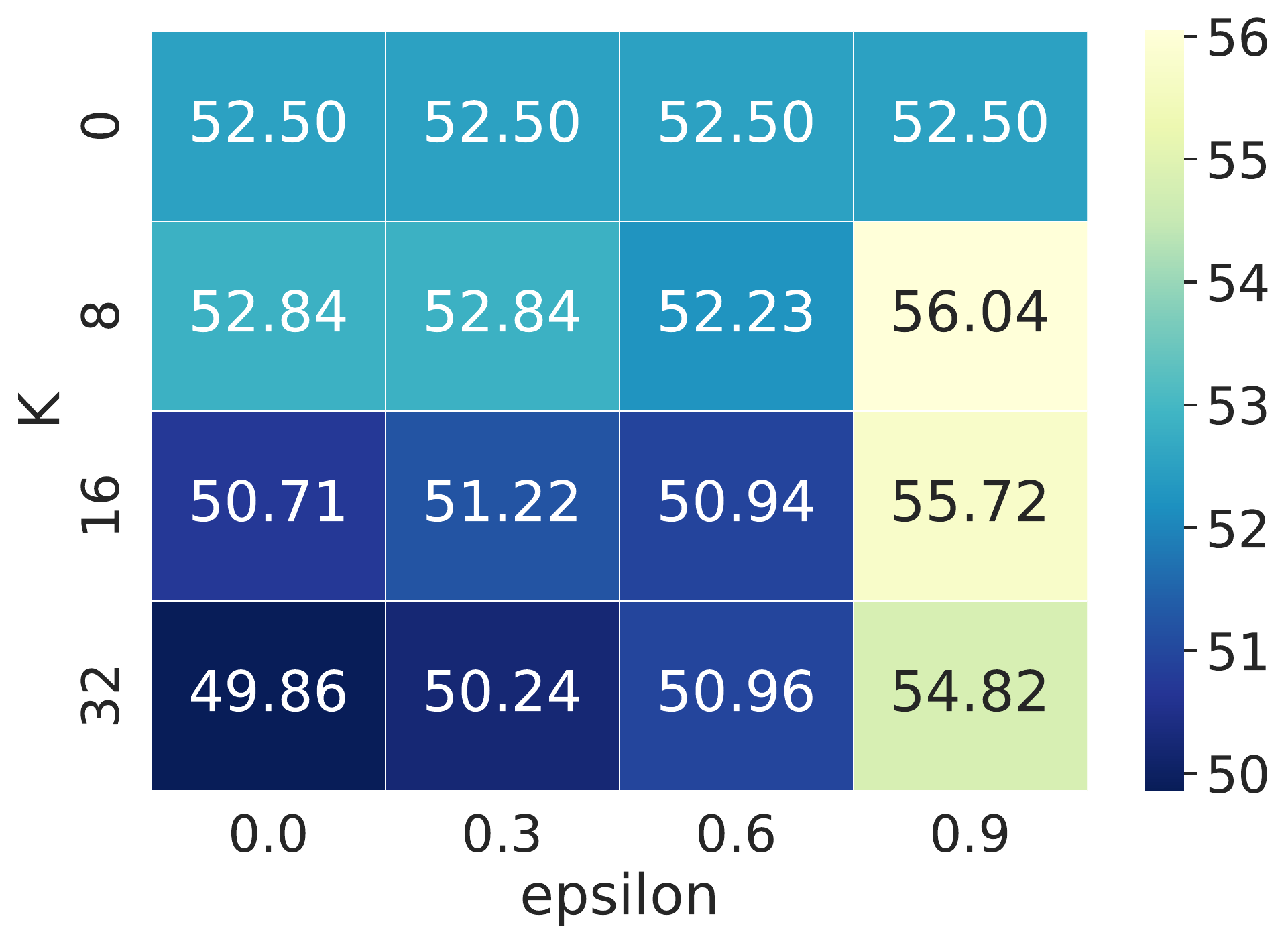}
    }
    \caption{Hyper-parameter Study of $K$ and $\epsilon$.}
    \label{fig:hyper_k_e_a}
\end{figure}

The results on dataset Liar and IMDB are shown in Figure~\ref{fig:hyper_k_e_a}.

\subsection{Influences of Batch Size $k_1$,$k_2$}

The results in Table V show that the proposed approach has stable improvements over different batch sizes. To be specific, SHGN with HDHGR decreases little when decreasing the batch size. We can adjust it to balance efficiency and effectiveness.

\begin{table}[t]
    \centering
    \caption{Impact of Batch Size $k_1,k_2.$}
    \resizebox{.9\linewidth}{!}{
    \begin{tabular}{ccc|cc}
    \toprule
       \multirow{2}{*}{Batch Size}  & \multicolumn{2}{c}{Actor} & \multicolumn{2}{c}{Liar} \\
       & Macro-F1 & Micro-F1 & Macro-F1 & Micro-F1 \\
    \midrule
        $1000\times 1000$ & $69.79\pm 0.51$ & $78.58\pm 0.32$ & $20.34\pm 0.65$ &  $24.31\pm 0.53$\\
        $5000\times 5000$ & $70.27\pm 0.88$ & $79.44\pm 0.55$ & $20.00\pm 0.51$ & $23.98\pm 0.40$\\
        $10000\times 10000$ & $71.26\pm 0.77$ & $80.00\pm 0.31$ & $21.23\pm 0.33$ & $25.32\pm 0.22$\\
        $N_\tau\times N_\tau$ & $72.73\pm 0.35$ & $81.36\pm 0.37$ & $22.29\pm 0.72$ & $23.82\pm 1.08$\\
    \bottomrule
    \end{tabular}
    }
    \label{tab:batch_size}
\end{table}

\end{document}